\documentclass[manuscript]{acmart}
\AtBeginDocument{%
  }

\usepackage{makecell}

\begin{document}

\title{Bringing Religious Values Into Design: How Theologians and Engineers See the Same Patterns Differently}

\author{Louisa Conwill}
\email{lconwill@nd.edu}
\orcid{0009-0001-7116-266X}
\affiliation{%
  \institution{Department of Computer Science and Engineering}
  \institution{University of Notre Dame}
  \city{Notre Dame}
  \state{Indiana}
  \country{USA}
}

\author{Megan Levis Scheirer}
\email{mlevis@nd.edu}
\orcid{0000-0002-3328-2732}
\affiliation{%
  \institution{Institute for Social Concerns}
  \institution{University of Notre Dame}
  \city{Notre Dame}
  \state{Indiana}
  \country{USA} \\
  \institution{Department of Computer Science and Engineering}
  \institution{University of Notre Dame}
  \city{Notre Dame}
  \state{Indiana}
  \country{USA}
}

\author{Karla Badillo-Urquiola}
\email{kbadill3@nd.edu}
\orcid{0000-0002-1165-3619}
\affiliation{%
  \institution{Department of Computer Science and Engineering}
  \institution{University of Notre Dame}
  \city{Notre Dame}
  \state{Indiana}
  \country{USA}
}

\author{Walter J. Scheirer}
\email{wscheire@nd.edu}
\orcid{0000-0001-9649-8074}
\affiliation{%
  \institution{Department of Computer Science and Engineering}
  \institution{University of Notre Dame}
  \city{Notre Dame}
  \state{Indiana}
  \country{USA}
}

\renewcommand{\shortauthors}{Conwill et al.}

\begin{abstract}
Religious and spiritual (R/S) principles are having an increasing influence on human-centered technology design in both academia and industry. These efforts may provoke questions of reflexivity especially when designing for principles one may be newly exposed to. We explore how different perspectives may impact designing for R/S values by interviewing eight software engineers with little religious affiliation and eight theologians with academic expertise in \textit{Catholic Social Teaching} (CST) -- the theological tradition underpinning Pope Leo XIV’s AI encyclical \textit{Magnifica Humanitas} -- about how principles of CST translate to technology designs. We found that software engineers and theologians frequently responded similarly. The differences in their responses suggested that while software engineers did not misunderstand the principles, theologians had a broader understanding. Our results suggest that differences in both experience with values and professional domain can impact how values are translated into designs.
\end{abstract}



\maketitle

\section{Introduction}

Since ancient times, people have looked to religion and spirituality (R/S) for guidance on how to live a good and meaningful life. The majority of the world's major R/S traditions have developed over the past hundreds and even thousands of years, making them treasure-troves of time-tested wisdom. As technology plays an increasing role in our lives, R/S traditions have grappled with how to live with technology well. To offer just a few of many examples, the Amish have notably exercised careful discernment in adopting technologies~\cite{ems2026amish}, Jews traditionally limit technology use on Shabbat~\cite{myjewishlearning_electricity_shabbat}, Catholic Social Teaching was developed in the Industrial Revolution era to respond to concerns with technology~\cite{compendium}, and mindfulness Buddhism can help foster a healthier relationship with attention economy technologies such as social media~\cite{gassho2026socialmedia}.

In the academy, R/S research in HCI began in 1996~\cite{buie2013spirituality} and has been progressively growing in the past 10-15 years~\cite{wolf2024still, kitson2025diving, wolf2026nospirituality}. R/S researchers in HCI argue that because religion and spirituality are deeply embedded in the lives of people around the world, HCI as a field should pay attention to the ways that R/S shapes our relationship with technology~\cite{wolf2026nospirituality}. Furthermore, a serious consideration of religious ethical understandings (especially, but not limited to, non-western religions) has been proposed as an antidote to the hegemonic western \emph{secular} notions of ethics that dominate technology ethics conversations~\cite{ahmed2022situating}.

More recently there has been interest from the tech industry in applying religious understandings of ethics to technology products, especially to AI. In the 2020s, the Catholic Church partnered with companies such as Microsoft and IBM in a commitment to AI ethics~\cite{romecall}, advised multiple Silicon Valley leaders~\cite{germain2023}, and published the encyclical \textit{Magnifica Humanitas}~\cite{magnificaHumanitas} on AI ethics, inviting Anthropic co-founder Chris Olah to speak at the document's release~\cite{anthropicmagnifica}. The Catholic Church isn't the only faith tradition shaping AI ethics: Anthropic has also met with Protestant Christian leaders~\cite{tiku2026can}, and both Anthropic and OpenAI participated in a Faith-AI Covenant roundtable with Hindu, Sikh, Baha'i, Greek Orthodox, and Latter-Day Saints leaders to determine universal ethical standards~\cite{fauria2026openai}. There has also been a recent increase in initiatives for bringing religious perspectives into technology ethics, attempting to drive real change by bridging the gap between academia, industry, and policy. These include interdisciplinary networks such as the AI and Faith organization~\cite{aiandfaith}, the Faith Family and Technology Network~\cite{fftn}, and the Buddhism and AI Initiative~\cite{buddhismAndAI}, as well as technical efforts to develop faith-informed benchmarks for LLMs~\cite{wingate2026omissive, kadous2026jaleesbench}.

As religious values are increasingly drawn upon for practical guidance in implementation, a question of \textit{reflexivity}, the ability to identify the impact of one's personal background and values on one's own analysis or interpretation, arises. When designing for particular values, a lack of reflexivity could mean the chosen values have a different influence on the design than intended~\cite{timmermans2014reflexivity}. For example, if a Christian engineer at Anthropic were asked to design an AI system that abides by Sikh ethical principles, would this engineer be able to correctly interpret the Sikh principle well enough to design for it? Further, because developers often lack resources for ethical guidance and training~\cite{griffin2025ethical}, can we rely on developers alone to implement religious values in their systems? Is it necessary to bring value experts such as religious leaders into the design process in these cases?

In this work, we compare understandings of how religious principles translate into software designs between two groups: theologians (academic religious value experts) with limited experience in computer science and software developers with no strong religious affiliation. To guide our work, we posed the following research questions:
\begin{itemize}
    \item \textbf{RQ1:} How do software engineers and theologians understand the designs of social media and LLMs as embodying religious principles? Where do their understandings differ?
    \item \textbf{RQ2:} How do we best leverage the perspectives of software engineers and theologians in designing for R/S values?
\end{itemize}

We approach these research questions through investigating the particular case of designing for the principles of Catholic Social Teaching (CST): the theological tradition underpinning Pope Leo XIV's AI Encyclical \textit{Magnifica Humanitas}. We conducted semi-structured interviews with software developer and theologian participant groups, asking participants to identify software design patterns that they felt embodied particular principles of CST and justify their choices. We answer \textbf{RQ1} by comparing both the design patterns the members of each group selected and the reflexive thematic analysis of their justifications. We answer \textbf{RQ2} by reflecting on the differences in results between groups. We found that in many cases theologians and software engineers responded similarly, suggesting that software engineers are able to quickly grasp unfamiliar principles. The few cases of differences in responses revealed that the software engineers didn't necessarily have misunderstandings, but the theologians had a broader understanding of the CST principles. Our results suggested two distinct causes for these differences in responses: a \textit{reflexivity of experience}, relating to differences in expertise with particular values, and a \textit{reflexivity of domain}, relating to how one's professional discipline impacts one's way of thinking.

The contributions of this work are:
\begin{itemize}
    \item The first empirical study to our knowledge investigating reflexivity in the context of designing for religious values
    \item The empirical contributions of our interviews, which provide insight into how software engineers and theologians interpret the principles of CST in the context of today's digital technologies, and their differences in understanding
    \item The characterization of these differences as \textit{reflexivity of experience} and \textit{reflexivity of domain}, which may apply in contexts of designing for other R/S values
    \item A discussion of how our results can inform how to effectively design for CST and other R/S values in real-world settings
\end{itemize}

As more technology companies seek to integrate religious values into design, our findings surface considerations about how form design teams that most effectively capture these values in technology designs.

\section{Related Work}
We situate our work at the intersection of HCI R/S research, focusing on studies that engage the perspectives of religious experts and employ value sensitive design and reflexivity. We highlight the unique contributions of our work.

\subsection{History and Trajectory of Religion and Spirituality Research in HCI}
Religion and spirituality-related research is a growing area within HCI. In a seminal work from 2013, Buie and Blythe found that ``techno-spirituality'' research had been growing since 1996 but still relatively lagged behind the production of R/S-related software applications being developed in the wild~\cite{buie2013spirituality}. As a notable step forward, R/S HCI scholars hosted a CHI 2022 workshop focused on integrating faith, spirituality, and religion in HCI, and published key ideas from the workshop in a special issue of ACM Interactions magazine~\cite{rifat2022integrating}. More recent reviews  led by Sara Wolf~\cite{wolf2024still, wolf2026nospirituality} and Åhman~\cite{ahman_Thoren_2025} suggest that while R/S research in HCI is growing in volume and diversity, the field still remains underdeveloped. The most recent survey, published by Wolf et al. in 2026, found that design-oriented work was the most common form of engagement in R/S research, including technologies explicitly designed to evoke R/S experiences (e.g., virtue reality that enables transcendent experiences) and technologies that support R/S practices without evoking R/S experiences (e.g., apps that help Muslim women align menstrual tracking with religious observance) ~\cite{wolf2026nospirituality}. R/S research conducted after this survey has continued in these primary areas, including a study of how algorithmic experiences of spiritual content support or hinder spiritual growth in the Islamic context~\cite{jahan2026progress} and a study of techno-spiritual practices of Black Christian young adults~\cite{smith2026blackchristian}. Other recent activities demonstrate growing engagement in R/S and HCI, including R/S related workshops and panels at major conferences~\cite{kitson2025diving, markum2024navigating, markum2023designing, smith2024undesigning} and the establishment of the SPIRITED Collective research network~\cite{spiritedhci}. 

\subsection{Value Sensitive Design and Reflexivity}
Value sensitive design (VSD) is a foundational design method for integrating human values into the technology design process~\cite{friedman2019value}, and is one approach for designing for R/S concerns~\cite{smith2026spirit}. While VSD's methodologies are varied, they all take a tripartite approach integrating conceptual, technical, and empirical inquiries~\cite{friedman2019value}. The conceptual inquiry aims to determine and define the values to design for in the context of the chosen technology. This process can be aided by the empirical inquiry, which entails research studies to gain understanding about the users' values and practices. Finally, the technical inquiry entails implementing the values in technical systems. VSD acknowledges that different people may come up with different definitions of values in the conceptual inquiry~\cite{friedman2013value}, surfacing questions of reflexivity in this stage. In one specific example, Alsheik et al. found that when taking a VSD approach, cross-cultural interpretations of a particular value between Western and Islamic cultures led to vastly different design implications~\cite{alsheikh2011whose}. More broadly, Borning and Muller recommend making the voice and values of the researchers more visible in VSD-related research~\cite{borning2012next}. Timmermans and Mittelstadt propose paths forward for increased culturally-sensitive reflexivity in VSD including visibility of the VSD practitioner's personal background and investigating the cultural bias of existing value frameworks popular among VSD practitioners~\cite{timmermans2014reflexivity}. In this work, we investigate how value experts (theologians) and value practitioners (software engineers) differ in their conceptual understandings of religious principles within a technical context. We do this to better understand more clearly how value translation differs between these two groups specifically in the context of religious values. While transparency of one's personal background is an important starting point for reflexivity, we aim to investigate in a more targeted way how different understandings of values may impact the resulting designs.

\subsection{Designing for R/S Values}
In addition to R/S research that proposes how technology design can support particular spiritual experiences or religious practices, there is also a growing body of research that considers how religious principles can guide the design of everyday technologies that do not have a direct spiritual or religious objective. The 2022 special issue of ACM Interactions magazine focusing on R/S research included a number of such articles, including how designing for Christian values can help resist attention economy designs~\cite{hiniker2022reclaiming} and how designing for Jewish values can promote healthier online speech~\cite{hammer2022individual}. Beyond the special issue, other HCI research has considered how Islamic values can inform LLM alignment~\cite{campbell2026codesigning} or technology design more broadly~\cite{noor2026decoding}. Conwill et al. put forth a design method for cataloging design patterns that abide by particular religious or philosophical principles and used this method to generate social media designs that embody the principles of Catholic Social Teaching (CST)~\cite{conwill2025design}. Lad et al. built on this study, using Conwill et al.'s method to generate design patterns for LLMs that embody CST~\cite{lad2026possible}. We use the design patterns documented by Conwill et al. and Lad et al. in our interviews as probes for understanding how our participants translate religious principles into designs.

\subsubsection{Integrating Faith Expert Perspectives in Design}
Some such works consult faith experts, particularly chaplains or spiritual care providers, in order to align with religious or spiritual aims in design. In the majority of the studies we found, the chaplains' opinions were solicited for the purpose of designing technologies that support some kind of spiritual care. For example, Bezabih et al. interviewed chaplains to understand how spiritual care can or cannot be expanded to online spaces~\cite{bezabih2025meeting}. Smith et al. consulted the perspectives of spiritual care providers to inform a design framework for delivering spiritual care using digital technologies~\cite{smith2026spirit}. Wester et al. consulted chaplains to understand how to design AI chatbots for conversational care~\cite{wester2026chaplains}.

In contrast to these prior works, our work solicits the perspectives of faith experts for their understanding of how ethical values translate to technology. The technology designs we consider are not necessarily technologies for an explicit spiritual or religious experience. Rather, we seek the opinions of faith experts to understand their perspectives on how religious values could inform the designs of more everyday technologies such as social media and AI. This is more similar to the work of Lutz et al., which interviewed religious people, including but not limited to religious experts, to better understand how to avoid unintended harms toward religious people that are embedded in technologies, and to propose design values inspired by religious perspectives~\cite{lutz2025goodfaith}. Another work by Lutz and Weyl engaged faith experts for LLM red teaming efforts to expand understandings of digital safety~\cite{lutz2026redteaming}. These two papers led by Lutz studied how the perspectives of religious experts directly impact technology designs. In contrast, our work is concerned with the reflexivity of religious experts and non-experts: if different backgrounds or worldviews lead to different interpretations of religiously-inspired design principles. 

\section{Method}
We evaluated how background and expertise informs designing for religious principles by comparing theologians and software engineers in a value translation task. The theologians (academics studying religious belief from a religious perspective) had strong familiarity with the chosen principles and the software engineers had both little familiarity with these particular principles and little to no religious affiliation in general.

We situate our methodology within the conceptual, technical, and empirical investigations of VSD. While traditional VSD often employs open-ended co-design to elicit stakeholder values, our study focuses specifically on the value translation process. We assessed such understandings through semi-structured interviews (empirical). Participants were presented with six religious principles as well as design patterns -- formally documented descriptions of software designs~\cite{gamma1995pattern} -- for social media and AI platforms. For each religious principle, participants were asked to select which design patterns they would choose if they were to design a system to abide by that principle (technical) and to justify why the chosen design patterns embody that principle (conceptual).

Because participants were given a list of design patterns to select from rather than envisioning their own designs, as would occur in a traditional VSD co-design workshop, responses do not reflect the entirety of each participants' understanding of how a particular religious principle would translate to software designs. Rather our matching task functions as a diagnostic probe. By using standardized design patterns as fixed empirical targets, we isolate how differences in domain expertise (software engineering vs. theology) shape a designer’s capacity for value translation.

The six religious principles we used in our study were six of the main principles of Catholic Social Teaching (CST): the Catholic Church's doctrine on human dignity and societal good that has a particular focus on how to live well with industrialization and technology~\cite{conwill2024virtue}. Specifically, the six principles were \textit{life and dignity of the human person; the call to family, community, and participation; option for the poor and vulnerable; solidarity; subsidiarity;} and \textit{care of God's creation.} Definitions of these principles can be found in the appendix. We chose CST because of the recent excitement that Pope Leo's AI encyclical \textit{Magnifica Humanitas} spurred about CST beyond the Catholic world~\cite{fauria2026pope}. This publication applies CST to its analysis of AI and advances the doctrine of CST through its teachings. As interest increases among both Catholics and non-Catholics to practically apply the teachings of \textit{Magnifica Humanitas} to technology designs, it is particularly timely to study reflexivity in the context of CST. 

There are also methodological conveniences to studying CST in comparison to other traditions. Primarily, we were able to re-use the existing social media and LLM design patterns by Conwill et al.~\cite{conwill2024virtue} and Lad et al.~\cite{lad2026possible} that were cataloged to embody the principles of CST using a VSD-inspired methodology~\cite{conwill2024virtue}. Although participants did not know which principles the design patterns were intended to embody, using CST-inspired patterns against CST principles made them better probes than using arbitrary patterns (or patterns inspired by a different religious tradition). This shared thematic foundation increased the likelihood that participants would connect patterns to CST principles (as opposed to seeing no connections), generating sufficient data for a meaningful comparison. Secondarily, CST was a convenient value system to work with because as computer scientists at a Catholic university it was easier to recruit Catholic theologians than theologians of any other religion. (However, not all theologians who we interviewed were affiliated with our university.)

\subsection{Participant Recruitment}
We recruited eight theologians and eight software engineers using word of mouth and then snowball sampling~\cite{goodman1961snowball}. We aimed to recruit as many participants as we feasibly could while also maintaining equal numbers of participants per group for a balanced comparison. Our reflexive thematic analysis results suggested we reached data saturation~\cite{saunders2018saturation} with this number of participants. The theologians were required to have PhDs in theology with CST as a primary part of their academic focus and to also have little expertise in computer science or software engineering. The engineers were required to be currently working in the technology industry in a software engineering or applied research scientist role and self-identify as having no strong religious identity or affiliation. The software engineers were also required to have little to no familiarity with CST. The majority of participants in both groups were United States citizens although both the theologian group and the software engineer group had at least one non US-citizen participant. All participants resided in the United States with the exception of one software engineer who was a US citizen living in Europe. Table \ref{tab:participant_demographics} summarizes our participant demographics.

\begin{table*}[htb]
\caption{\textbf{Participant Demographics}}
\label{tab:participant_demographics}
\begin{tabular}{|cllll|}
\hline
\textbf{ID\#} & \textbf{Gender} & \textbf{Race/Ethnicity} & \textbf{Age} & \textbf{Number of Years in Field}    \\ \Xhline{1.5pt}
SE1 &   Man  &   White/Hispanic          &   31  &   9    \\ \hline
SE2 &   Man  &   White/Middle Eastern & 30 & 8 \\ \hline
SE3 &   Woman & East Asian  & 32 & 2 industry, 8 academia \\ \hline
SE4 & Man & White & 34 & 7 industry, 5 academia \\ \hline
SE5 & Man & White & 29 & 2 industry, 5 academia \\ \hline
SE6 & Man & White/Hispanic & 32 & 10 \\ \hline
SE7 & Man & White/Latino & 33 & 6 industry, 4 academia \\ \hline
SE8 & Woman & South Asian & 31 & 9 \\ \Xhline{1.5pt}
T1 & Man & White & 43 & 17 \\ \hline
T2 & Man & White & 44 & 22 \\ \hline
T3 & Woman & White & 35 & 11 \\ \hline
T4 & Man & Latino & 43 & 10 \\ \hline
T5 & Man & White & 61 & 40 \\ \hline
T6 & Man & White & 52 & 31 \\ \hline
T7 & Man & White & 36 & 12 \\ \hline
T8 & Man & White & 44 & 14 \\ \hline
\end{tabular}
\end{table*}

During the course of our interviews one \emph{historian} with CST expertise was recruited for the theologian group by mistake. Her design pattern selections and justifications are not included because she did not fit our exclusion criteria of having a PhD in \emph{theology}; however we include observations from her interview that we believe provide additional context for our findings.

\subsection{Interview Protocol}
The semi-structured interviews were conducted in person if feasible, otherwise on Zoom, FaceTime, or Google Meet with cameras turned on. The first author conducted the interviews. The interviews were recorded and auto-transcribed with Zoom or with the Apple Voice Memos app on the interviewer's iPhone (two Zoom interviews were uploaded to Otter.AI for audio transcription because the interviewer did not correctly save the Zoom transcription). One in-person participant who consented to the interview being recorded but not auto-transcribed was recorded using the Diarium app which takes audio recordings without auto-transcribing them (the interviewer was not able to turn off the auto-transcription feature on her Voice Memos app). The interviewer transcribed as much of this interview as possible by hand in real time, and consulted the audio recording after the interview to fill in missing gaps.

Participants were first asked to share how many years of work experience they have in their field (post-undergraduate studies, but including graduate studies if applicable) and optionally share demographic information. Then the interviewer shared with participants a brief background on the motivation of the study, including real-world examples of technology companies consulting religious leaders for ethical advice.

\begin{figure}
    \centering
    \includegraphics[width=\linewidth]{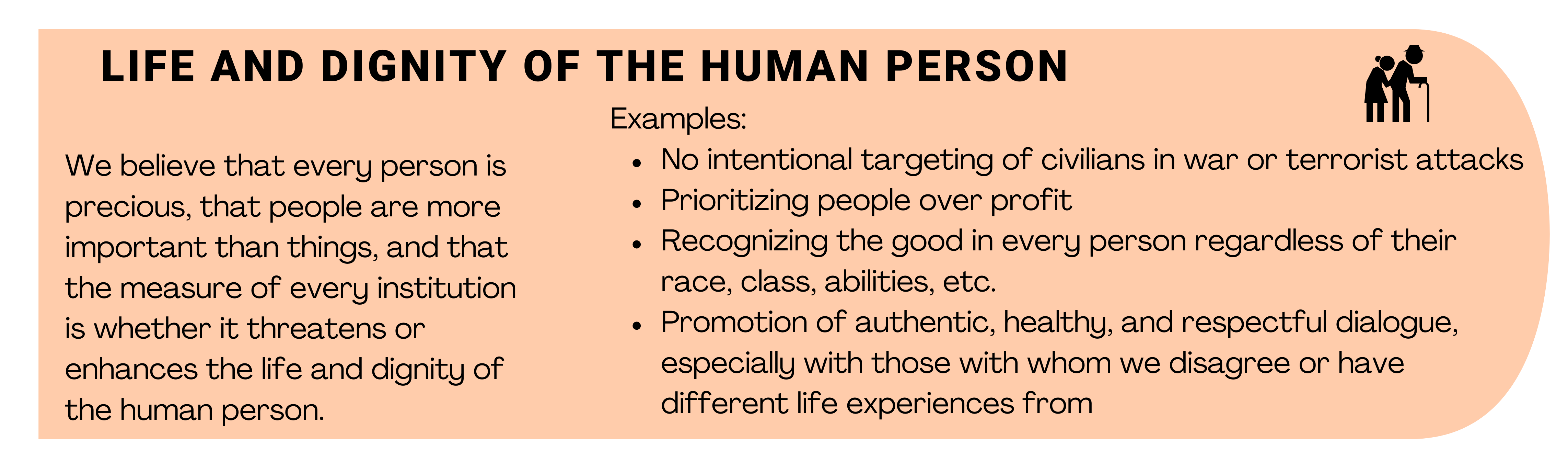}
    \caption{The explanation of the CST principle \textit{life and dignity of the human person} from the handout provided to interview participants. Each principle's explanation included a basic overview as well as examples of the principle in action outside of the context of software design. The full handout is in the appendix.}
    \label{fig:infographic-example}
    \Description{A text box with a peach-colored background. The title reads ``LIFE AND DIGNITY OF THE HUMAN PERSON'' and there is an icon of a hunched over woman holding the hand of a hunched over man with a cane. The box reads: we believe that every person is precious, that people are more important than things, and that the measure of every institution is whether it threatens or enhances the life and dignity of the human person. Examples:
    \begin{itemize}
        \item No intentional targeting of civilians in war or terrorist attacks
        \item Prioritizing people over profit
        \item Recognizing the good in every person regardless of their race, class, abilities, etc.
        \item Promotion of authentic, healthy, respectful dialogue, especially with those with whom we disagree or have different life experiences from
    \end{itemize}}
\end{figure}

\begin{figure}
    \centering
    \includegraphics[width=\linewidth]{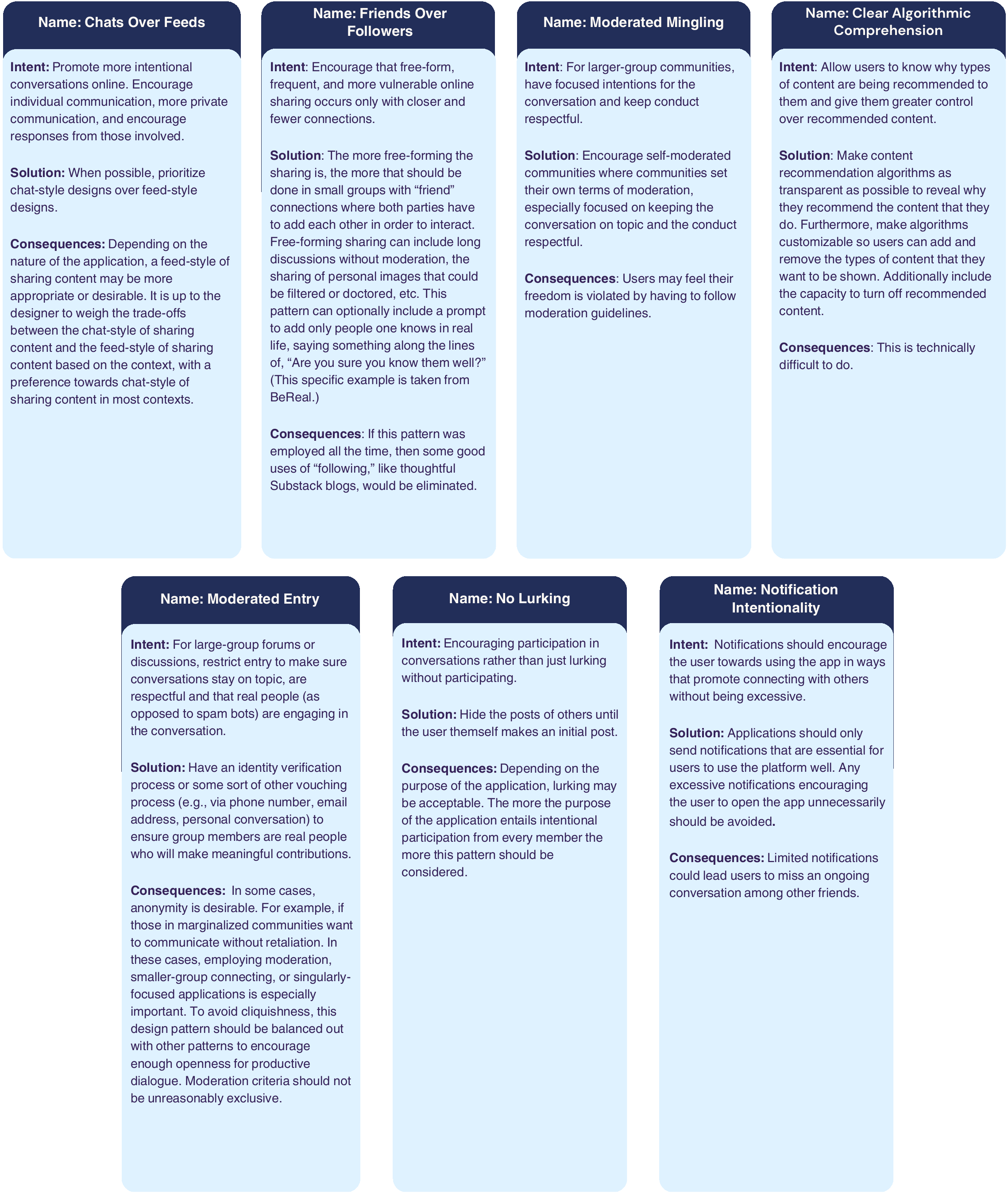}
    \caption{Social media design patterns documented by Conwill et al.~\cite{conwill2025design} that embody one or more of the six principles of Catholic Social Teaching. These design patterns were used as probes in the interviews to assess how software engineers and theologians understand how religious principles would translate to technology designs.}
    \label{fig:social-media-design-patterns}
    \Description{The seven design patterns for social media are listed with their name, intent, solution, and consequences. They are as follows. Pattern 1. Name: Chats Over Feeds. Intent: Promote more intentional conversations online. Encourage individual communication, more private communication, and encourage responses from those involved. Solution: When possible, prioritize chat-style designs over feed-style designs. Consequences: Depending on the nature of the application, a feed style of sharing content may be more appropriate or desirable. It is up to the designer to weigh the trade-offs between the chat-style of sharing content and the feed-style of sharing content based on the context, with a preference towards chat-style of sharing content in most contexts. Pattern 2. Name: Friends Over Followers. Intent: Encourage that free-form, frequent, and more vulnerable online sharing occur only with closer and fewer connections. Solution: The more free-forming the sharing is, the more that should be done in small groups with “friend” connections where both parties have to add each other in order to interact. Free-forming sharing can include long discussions without moderation, the sharing of personal images that could be filtered or doctored, etc. This pattern can optionally include a prompt to add only people one knows in real life, saying something along the lines of, “Are you sure you know them well?” (This specific example is taken from BeReal.) Consequences: If this pattern was employed all the time, then some good uses of “following,” like thoughtful Substack blogs, would be eliminated. Pattern 3. Name: Moderated Mingling. Intent: For larger-group communities, have focused intentions for the conversation and keep conduct respectful. Solution: Encourage self-moderated communities where communities set their own terms of moderation, especially focused on keeping the conversation on topic and the conduct respectful. Consequences: Users may feel their freedom is violated by having to follow moderation guidelines. Pattern 4. Name: Clear Algorithmic Comprehension. Intent: Allow users to know why types of content are being recommended to them and give them greater control over recommended content. Solution: Make content recommendation algorithms as transparent as possible to reveal why they recommend the content that they do. Furthermore, make algorithms customizable so users can add and remove the types of content that they want to be shown. Additionally include the capacity to turn off recommended content. Consequences: This is technically difficult to do. Pattern 5. Name: Moderated Entry. Intent: For large-group forums or discussions, restrict entry to make sure conversations stay on topic, are respectful and that real people (as opposed to spam bots) are engaging in the conversation Solution: Have an identity verification process or some sort of other vouching process (e.g. via phone number, email address, personal conversation, etc.) to ensure group members are real people who will make meaningful contributions. Consequences: In some cases, anonymity is desirable. For example, if those in marginalized communities want to communicate without retaliation. In these cases, employing moderation, smaller-group connecting, or singularly-focused applications are especially important. To avoid cliquishness, this design pattern should be balanced out with other patterns to encourage enough openness for productive dialogue. Moderation criteria should not be unreasonably exclusive. Pattern 6. Name: No Lurking. Intent: Encouraging participation in conversations rather than just lurking without participating. Solution: Hide the posts of others until the user themselves makes an initial post. Consequences: Depending on the purpose of the application, lurking may be acceptable. The more the purpose of the application entails intentional participation from every member the more this pattern should be considered. Pattern 7. Name: Notification Intentionality. Intent:  Notifications should encourage the user towards using the app in ways that promote connecting with others without being excessive. Solution: Applications should only send notifications that are essential for users to use the platform well. Any excessive notifications encouraging the user to open the app unnecessarily should be avoided. Consequences: Limited notifications could lead users to miss an ongoing conversation among other friends.}
\end{figure}

\begin{figure}
    \centering
    \includegraphics[width=\linewidth]{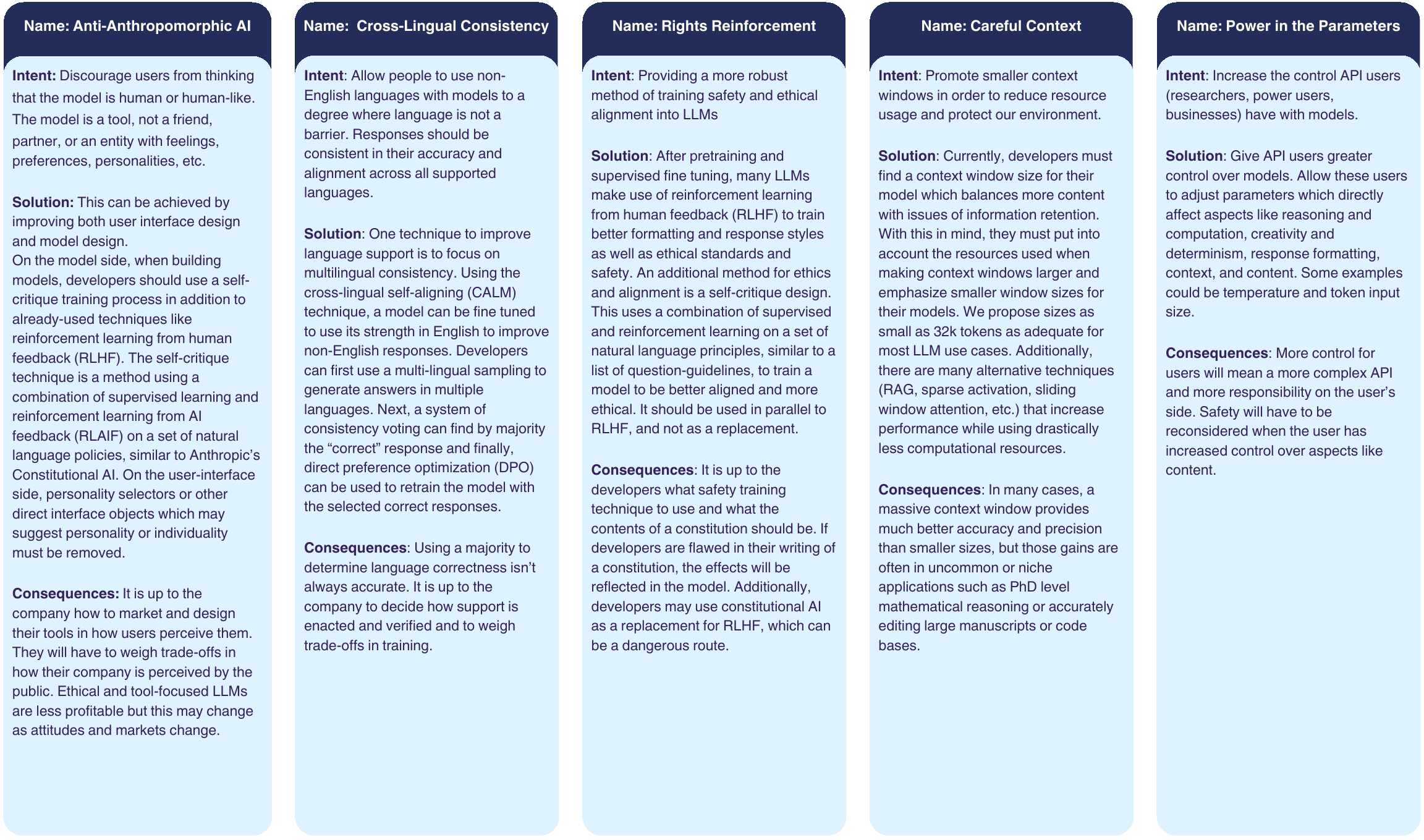}
    \caption{Large language model design patterns documented by Lad et al.~\cite{lad2026possible} that embody one or more of the six principles of Catholic Social Teaching. These design patterns were used as probes in the interviews to assess how software engineers and theologians understand how religious principles would translate to technology designs.}
    \label{fig:llm-design-patterns}
    \Description{The five design patterns for large language models are listed with their name, intent, solution, and consequences. They are as follows. Pattern 1. Name: Anti-Anthropomorphic AI. Intent: Discourage users from thinking that the model is human or human-like. The model is a tool, not a friend, partner, or an entity with feelings, preferences, personalities, etc. Solution: This can be achieved by improving both user interface design and model design. On the model side, when building models, developers should use a self-critique training process in addition to already-used techniques like reinforcement learning from human feedback (RLHF). The self-critique technique is a method using a combination of supervised learning and reinforcement learning from AI feedback (RLAIF) on a set of natural language policies, similar to Anthropic’s Constitutional AI. On the user-interface side, personality selectors or other direct interface objects which may suggest personality or individuality must be removed. Consequences: It is up to the company how to market and design their tools in how users perceive them. They will have to weigh trade-offs in how their company is perceived by the public. Ethical and tool-focused LLMs are less profitable but this may change as attitudes and markets change. Pattern 2. Name:  Cross-Lingual Consistency. Intent: Allow people to use non-English languages with models to a degree where language is not a barrier. Responses should be consistent in their accuracy and alignment across all supported languages. Solution: One technique to improve language support is to focus on multilingual consistency. Using the cross-lingual self-aligning (CALM) technique, a model can be fine tuned to use its strength in English to improve non-English responses. Developers can first use a multi-lingual sampling to generate answers in multiple languages. Next, a system of consistency voting can find by majority the “correct” response and finally, direct preference optimization (DPO) can be used to retrain the model with the selected correct responses. Consequences: Using a majority to determine language correctness isn’t always accurate. It is up to the company to decide how support is enacted and verified and to weigh trade-offs in training. Pattern 3. Name: Rights Reinforcement. Intent: Providing a more robust method of training safety and ethical alignment into LLMs. Solution: After pretraining and supervised fine tuning, many LLMs make use of reinforcement learning from human feedback (RLHF) to train better formatting and response styles as well as ethical standards and safety. An additional method for ethics and alignment is a self-critique design. This uses a combination of supervised and reinforcement learning on a set of natural language principles, similar to a list of question-guidelines, to train a model to be better aligned and more ethical. It should be used in parallel to RLHF, and not as a replacement. Consequences: It is up to the developers what safety training technique to use and what the contents of a constitution should be. If developers are flawed in their writing of a constitution, the effects will be reflected in the model. Additionally, developers may use constitutional AI as a replacement for RLHF, which can be a dangerous route. Pattern 4. Name: Careful Context. Intent: Promote smaller context windows in order to reduce resource usage and protect our environment. Solution: Currently, developers must find a context window size for their model which balances more content with issues of information retention. With this in mind, they must put into account the resources used when making context windows larger and emphasize smaller window sizes for their models. We propose sizes as small as 32k tokens as adequate for most LLM use cases. Additionally, there are many alternative techniques (RAG, sparse activation, sliding window attention, etc.) that increase performance while using drastically less computational resources. Consequences: In many cases, a massive context window provides much better accuracy and precision than smaller sizes, but those gains are often in uncommon or niche applications such as PhD level mathematical reasoning or accurately editing large manuscripts or code bases. Name: Power in the Parameters. Intent: Increase the control API users (researchers, power users, businesses) have with models. Solution: Give API users greater control over models. Allow these users to adjust parameters which directly affect aspects like reasoning and computation, creativity and determinism, response formatting, context, and content. Some examples could be temperature and token input size. Consequences: More control for users will mean a more complex API and more responsibility on the user’s side. Safety will have to be reconsidered when the user has increased control over aspects like content.}
\end{figure}

Participants were provided with a CST principles handout (see figure \ref{fig:infographic-example}), a handout of Conwill et al.'s seven social media design patterns~\cite{conwill2025design} (see figure \ref{fig:social-media-design-patterns}) and a handout of Lad et al.'s five LLM design patterns~\cite{lad2026possible} (see figure \ref{fig:llm-design-patterns}). In comparison to the versions of the patterns published in the original papers, we removed the ``CST principles embodied'' from the design pattern descriptions so as not to bias our participants. Participants were given an opportunity to ask the interviewer clarifying questions about the technical elements of the design patterns. Then, for each principle of CST the participants were asked: ``if you were building a [social media]/[LLM] system and wanted to design for [CST principle], which design patterns would you choose, if any, and why?'' Participants first did this exercise with the social media design patterns, and then with the LLM design patterns. The full interview protocol is in the appendix.

The CST principles handout contained a brief description of each principle, as well as non-technical examples of that principle in action. See figure \ref{fig:infographic-example} for an example of one principle from the handout; the full handout is in the appendix. Participants were allowed to ask clarifying questions about the principles in general; however the interviewer did not answer any questions about how the principle would be expressed in a technological context. Participants were allowed to consult the handout as well as rely on any preexisting knowledge they had about the principle (particularly relevant for the theologians with CST expertise).

\subsection{Data Analysis Approach and Positionality Statement}
The data was primarily analyzed qualitatively using reflexive thematic analysis~\cite{braun2019reflecting} and descriptive counts. The first and second authors who performed the analysis are both white Christian women, the third author is a Latina Christian woman, and the last author is a white Christian male. All authors are academics in computer science who additionally have extensive knowledge of CST.

To answer \textbf{RQ1} we investigated reflexivity in both the technical and conceptual inquiry stages. We began at VSD's technical inquiry stage, counting how many participants in the software engineer and theologian groups chose each design pattern as embodying a particular CST principle. We also calculated the average number of design patterns participants in each group chose for each principle of CST in both the social media and LLM contexts.

In addition to comparing the descriptive counts between groups, we performed exploratory Fishers Exact Tests~\cite{upton1992fisher} (the most appropriate statistical test for our small sample size of N=8 per group) to highlight statistically significant and other possible divergences between the groups. To maintain statistical rigor, we apply a standard threshold ($p < 0.05$). Differences that do not meet this threshold are interpreted as descriptive trends rather than confirmatory statistical claims. Following established exploratory screening protocols (e.g., Hosmer \& Lemeshow~\cite{hosmer2013applied}), we adopted an exploratory cutoff of $p <= 0.20$ to identify non-significant divergences warranting further investigation (we refer to this as ``possible difference'').

To investigate reflexivity in the conceptual inquiry stage, we used reflexive thematic analysis~\cite{braun2019reflecting} to analyze the justifications for selecting design patterns. The first author went through each transcript and pulled all quotes pertaining to justifications of why selected design patterns embody particular CST principles. The first author then separately coded the set of quotes pertaining to each CST principle. For each CST principle, she first familiarized herself with that principle's set of quotes by re-reading them multiple times. She then generated a set of initial codes for each CST principle. The first author and the second author (an advising author) then held iterative discussions to ensure the codes appropriately reflected the intention of each quote. The first and second authors then collaboratively generated themes to reflect patterns among the codes. The third and fourth authors then reviewed and approved these theme groupings. These resulting themes reflect the participants' understandings of how each CST principle is embodied by the design patterns. This maps to comparisons of reflexivity in VSD's conceptual inquiry stage: how people with different backgrounds define values in the context of software.

To better contextualize how the found differences between groups can inform designing for R/S values (\textbf{RQ2}), we synthesize the results into two findings sections. The first section presents findings from RQ1 that suggest a \textit{reflexivity of experience}: that differences in experience with a value system can impact how one performs in both the conceptual and technical inquiries. The second section presents findings from RQ1 as well as supporting observations from interviews that suggest a \textit{reflexivity of domain}: that one's professional training can impact how one performs in a technical inquiry.

\section{Findings Part 1: Reflexivity of Experience}
In this section we share the counts of how many participants selected a design pattern as relating to each principle of CST and the reflexive thematic analysis of their justifications. For both we highlight similarities and differences in responses between software engineers and theologians. In the case of both tasks, major differences consistently resulted from a higher proportion of theologians choosing a design pattern and mentioning a justification. The software engineers did not contradict the theologians; they simply had a more limited view, expressing a subset of what the theologians articulated. In most cases the strongest similarities in justifications between software engineers and theologians corresponded to information about the CST principle listed on the handout. Similarly, in most cases the strongest differences in justifications (i.e. justifications mentioned frequently by theologians and infrequently by software engineers) corresponded to dimensions of the CST principles not articulated on the handout. These findings suggest a \textit{reflexivity of experience}, that having greater experience with a value system can broaden conceptual and technical understandings when translating that value into software.

\subsection{Similarities and Differences in Design Pattern Selection}
\begin{figure}
    \centering
    \includegraphics[width=\linewidth]{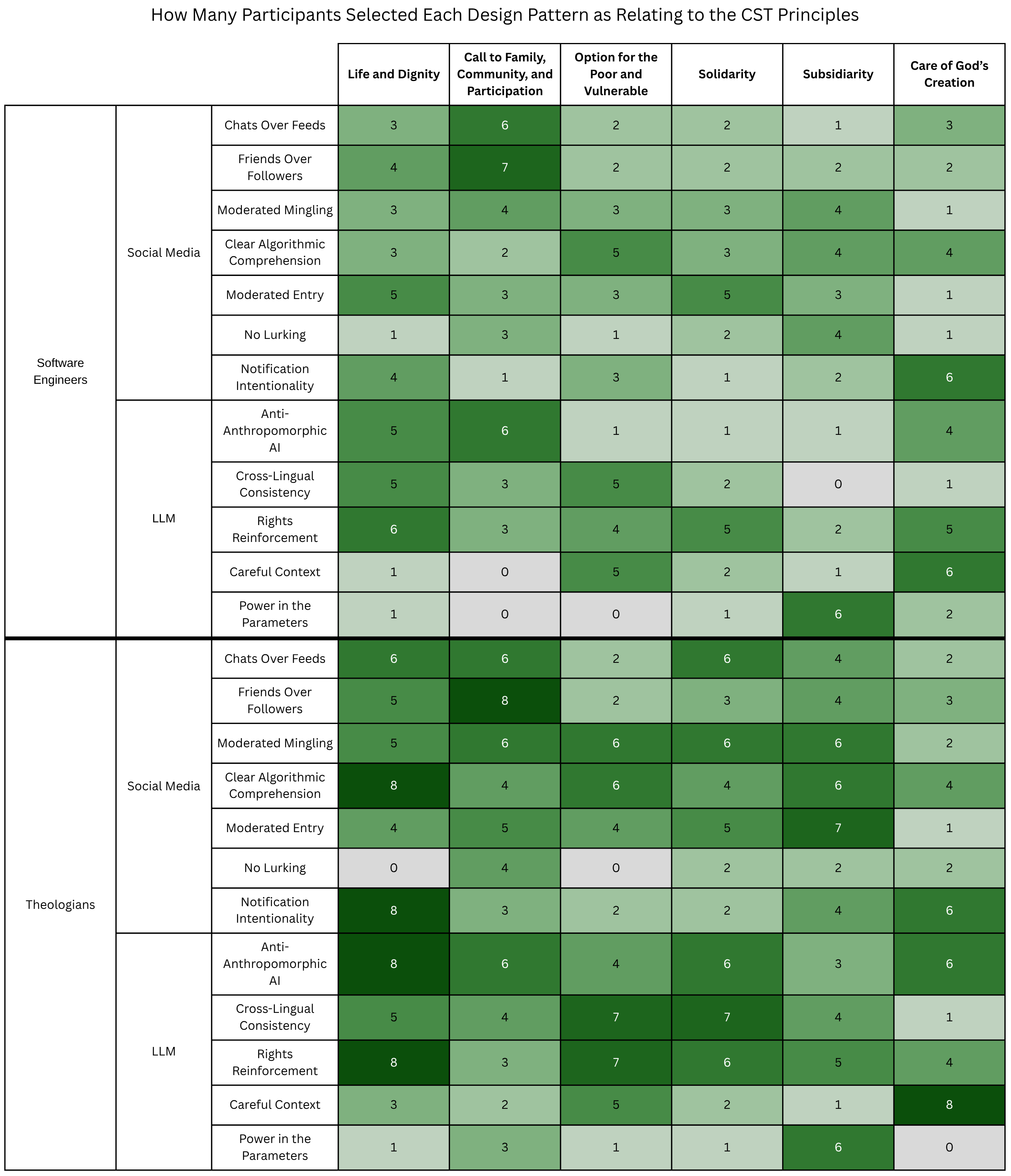}
    \caption{Heat map comparing how many participants per group thought a particular social media design pattern embodied each principle of Catholic Social Teaching. The biggest difference in responses appear to be for \textit{life and dignity of the human person}, \textit{option for the poor and vulnerable}, \textit{solidarity}, and \textit{subsidiarity}, with theologians selecting more design patterns as relating to these principles than software engineers.}
    \label{fig:design-patterns-selected}
    \Description{Heat map table titled "How Many Participants Selected Each Design Pattern as Relating to the CST Principles", comparing Software Engineers and Theologians.
    Data for Software Engineers:
    - Social Media:
    * Chats Over Feeds: Life and Dignity 3; Call to Family, Community, and Participation 6; Option for Poor and Vulnerable 2; Solidarity 2; Subsidiarity 1; Care of God's Creation 3.
    * Friends Over Followers: Life and Dignity 4; Call to Family, Community, and Participation 7; Option for Poor and Vulnerable 2; Solidarity 2; Subsidiarity 2; Care of God's Creation 2.
    * Moderated Mingling: Life and Dignity 3; Call to Family, Community, and Participation 4; Option for Poor and Vulnerable 3; Solidarity 3; Subsidiarity 4; Care of God's Creation 1.
    * Clear Algorithmic Comprehension: Life and Dignity 3; Call to Family, Community, and Participation 2; Option for Poor and Vulnerable 5; Solidarity 3; Subsidiarity 4; Care of God's Creation 4.
    * Moderated Entry: Life and Dignity 5; Call to Family, Community, and Participation 3; Option for Poor and Vulnerable 3; Solidarity 5; Subsidiarity 3; Care of God's Creation 1.
    * No Lurking: Life and Dignity 1; Call to Family, Community, and Participation 3; Option for Poor and Vulnerable 1; Solidarity 2; Subsidiarity 4; Care of God's Creation 1.
    * Notification Intentionality: Life and Dignity 4; Call to Family, Community, and Participation 1; Option for Poor and Vulnerable 3; Solidarity 1; Subsidiarity 2; Care of God's Creation 6.
    - LLM:
    * Anti-Anthropomorphic AI: Life and Dignity 5; Call to Family, Community, and Participation 6; Option for Poor and Vulnerable 1; Solidarity 1; Subsidiarity 1; Care of God's Creation 4.
    * Cross-Lingual Consistency: Life and Dignity 5; Call to Family, Community, and Participation 3; Option for Poor and Vulnerable 5; Solidarity 2; Subsidiarity 0; Care of God's Creation 1.
    * Rights Reinforcement: Life and Dignity 6; Call to Family, Community, and Participation 3; Option for Poor and Vulnerable 4; Solidarity 5; Subsidiarity 2; Care of God's Creation 5.
    * Careful Context: Life and Dignity 1; Call to Family, Community, and Participation 0; Option for Poor and Vulnerable 5; Solidarity 2; Subsidiarity 1; Care of God's Creation 6.
    * Power in the Parameters: Life and Dignity 1; Call to Family, Community, and Participation 0; Option for Poor and Vulnerable 0; Solidarity 1; Subsidiarity 6; Care of God's Creation 2.
    
    Data for Theologians:
    - Social Media:
    * Chats Over Feeds: Life and Dignity 6; Call to Family, Community, and Participation 6; Option for Poor and Vulnerable 2; Solidarity 6; Subsidiarity 4; Care of God's Creation 2.
    * Friends Over Followers: Life and Dignity 5; Call to Family, Community, and Participation 8; Option for Poor and Vulnerable 2; Solidarity 3; Subsidiarity 4; Care of God's Creation 3.
    * Moderated Mingling: Life and Dignity 5; Call to Family, Community, and Participation 6; Option for Poor and Vulnerable 6; Solidarity 6; Subsidiarity 6; Care of God's Creation 2.
    * Clear Algorithmic Comprehension: Life and Dignity 8; Call to Family, Community, and Participation 4; Option for Poor and Vulnerable 6; Solidarity 4; Subsidiarity 6; Care of God's Creation 4.
    * Moderated Entry: Life and Dignity 4; Call to Family, Community, and Participation 5; Option for Poor and Vulnerable 4; Solidarity 5; Subsidiarity 7; Care of God's Creation 1.
    * No Lurking: Life and Dignity 0; Call to Family, Community, and Participation 4; Option for Poor and Vulnerable 0; Solidarity 2; Subsidiarity 2; Care of God's Creation 2.
    * Notification Intentionality: Life and Dignity 8; Call to Family, Community, and Participation 3; Option for Poor and Vulnerable 2; Solidarity 2; Subsidiarity 4; Care of God's Creation 6.
    - LLM:
    * Anti-Anthropomorphic AI: Life and Dignity 8; Call to Family, Community, and Participation 6; Option for Poor and Vulnerable 4; Solidarity 6; Subsidiarity 3; Care of God's Creation 6.
    * Cross-Lingual Consistency: Life and Dignity 5; Call to Family, Community, and Participation 4; Option for Poor and Vulnerable 7; Solidarity 7; Subsidiarity 4; Care of God's Creation 1.
    * Rights Reinforcement: Life and Dignity 8; Call to Family, Community, and Participation 3; Option for Poor and Vulnerable 7; Solidarity 6; Subsidiarity 5; Care of God's Creation 4.
    * Careful Context: Life and Dignity 3; Call to Family, Community, and Participation 2; Option for Poor and Vulnerable 5; Solidarity 2; Subsidiarity 1; Care of God's Creation 8.
    * Power in the Parameters: Life and Dignity 1; Call to Family, Community, and Participation 3; Option for Poor and Vulnerable 1; Solidarity 1; Subsidiarity 6; Care of God's Creation 0.}
\end{figure}

Design pattern selections are reported as descriptive counts to illustrate observational trends between participant groups. Participant responses are shown in the heat map in figure \ref{fig:design-patterns-selected}. Significant or possible differences surfaced from the Fisher's Exact Test are mentioned in-line; the full Fisher's Exact Test results are in the appendix.

\subsubsection{General Trends}
There was no case in which the between-group difference in number of participants who selected a design pattern for a particular CST principle was greater than 5 participants. In cases when more software engineers chose a design pattern as applying to a CST principle than theologians did, it was only a difference of 1 or 2 more software engineers than theologians: the cases of bigger divergence (3 or more responses) always entailed more theologians selecting the design pattern.

\subsubsection{Life and Dignity of the Human Person}
The software engineers generally had mixed opinions about which social media design patterns related, with no response greater than N=5. In contrast, the theologians unanimously (N=8) agreed that clear algorithmic comprehension and notification intentionality embodied \textit{life and dignity}. This elicited significant difference on the Fisher's Exact Test for clear algorithmic comprehension, and possible difference for notification intentionality. Many theologians (N=6) also found chats over feeds to embody \textit{life and dignity}.

In the case of LLMs, both software engineers and theologians chose anti-anthropomorphic AI, cross-lingual consistency, and rights reinforcement more frequently than careful context or power in the parameters. Theologians had unanimity on anti-anthropomorphic and rights reinforcement. Although both groups had relatively high responses on anti-anthropomorphic AI relative to the other patterns, the Fisher's Exact Test suggested possible difference between groups for this pattern (N=5 for software engineers and N=8 for theologians).

\subsubsection{Call to Family, Community, and Participation}
Both the software engineer and theologian groups had higher numbers of responses for chats over feeds (N=6 for both groups) and friends over followers (N=7 for software engineers, N=8 for theologians). N=6 theologians thought moderated mingling applied while only N=4 software engineers did. The other social media patterns had relatively fewer responses for both groups, however the theologians had more responses for each pattern than the software engineers.

N=6 participants from both groups thought anti-anthropomorphic AI related to \textit{call to family}. The other LLM design patterns had fewer responses, however while N=0 software engineers thought careful context and power in the parameters related, N=2 theologians thought careful context related and N=3 theologians thought power in the parameters related. Power in the parameters elicited a possible difference on the Fisher's Exact Test.

\subsubsection{Option for the Poor and Vulnerable}
Software engineers (N=5) and theologians (N=6) agreed that clear algorithmic comprehension was the most relevant social media pattern to \textit{option for the poor and vulnerable}. The biggest difference was with moderated mingling: N=6 theologians thought moderated mingling applied whereas only N=3 software engineers did. For the other social media patterns, software engineers and theologians responded in the same numbers or with only a difference of one participant. Response numbers for the other patterns ranged from N=0 to N=4.

For the LLM patterns, relatively more participants per group thought cross-lingual consistency, rights reinforcement, and careful context applied, with more responses from the theologians than the software engineers in all cases. N=7 theologians selected cross-lingual consistency and rights reinforcement, in comparison to N=5 and N=4 software engineers, respectively. N=5 participants in both groups thought careful context applied. More theologians (N=4) thought anti-anthropomorphic applied than software engineers (N=1).

\subsubsection{Solidarity}
For social media, N=5 participants from each group thought moderated entry upheld \textit{solidarity}. For the software engineers this was the most agreed-upon pattern. In contrast, the theologians selected chats over feeds and moderated mingling more frequently (N=6 for both design patterns). These responses elicited possible difference on the Fisher's Exact Test for chats over feeds. 

For LLMs, N=5 software engineers and N=6 theologians thought rights reinforcement applied. For all other LLM design patterns, N=2 or fewer software engineers thought they related. However N=6 theologians thought anti-anthropomorphic related and N=7 theologians thought cross-lingual related. Both of these elicited significant difference on the Fisher's Exact Test.

\subsubsection{Subsidiarity}
For social media, no more than N=4 software engineers selected any design patterns as relating to subsidiarity. N=6 theologians thought moderated mingling and clear algorithmic comprehension related, and N=7 theologians thought moderated entry related. Moderated entry elicited possible difference on the Fisher's Exact Test.

For LLMs, N=6 participants from both groups thought power in the parameters related. While N=2 or fewer software engineers selected any of the other LLM patterns, N=3 theologians selected anti-anthropomorphic, N=4 selected cross-lingual consistency, and N=5 selected rights reinforcement. Cross-lingual consistency elicited possible difference on the Fisher's Exact Test.

\subsubsection{Care of God's Creation}
For social media, N=6 participants in both groups thought notification intentionality applied. The other design patterns had between N=1 and N=4 responses. Notably, every pattern had at most a difference of 1 response between groups.

For LLMs, in both groups more participants chose anti-anthropomorphic, rights reinforcement, and careful context as embodying \textit{care of God's creation}. In the case of careful context and anti-anthropomorphic, more theologians than software engineers chose the patterns (N=4 software engineers and N=6 theologians for anti-anthropomorphic; N=6 software engineers and N=8 theologians for careful context). For rights reinforcement, one more software engineer selected it than theologian (N=5 software engineers and N=4 theologians). 

\subsection{Similarities and Differences in Justifications}
\begin{figure}
    \centering
    \includegraphics[width=\linewidth]{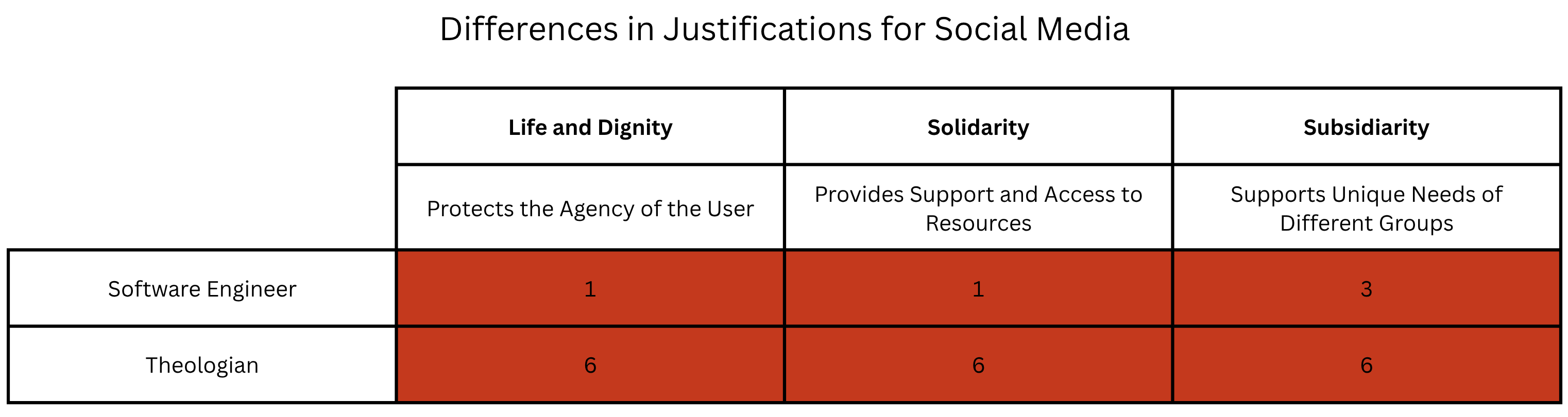}
    \caption{The themes that had the biggest difference between groups in number of participants who mentioned that theme for the social media design patterns. Any theme that had a between-group difference of 3 or more participants mentioning that theme in their justification is listed here.}
    \label{fig:sm-theme-differences}
    \Description{Table titled "Differences in Justifications for Social Media", comparing responses from Software Engineers and Theologians across three Catholic Social Teaching principles and their associated themes.
    - Life and Dignity (Protects the Agency of the User): Software Engineer: 1, Theologian: 6.
    - Solidarity (Provides Support and Access to Resources): Software Engineer: 1, Theologian: 6.
    - Subsidiarity (Supports Unique Needs of Different Groups): Software Engineer: 3, Theologian: 6.}
\end{figure}

\begin{figure}
    \centering
    \includegraphics[width=\linewidth]{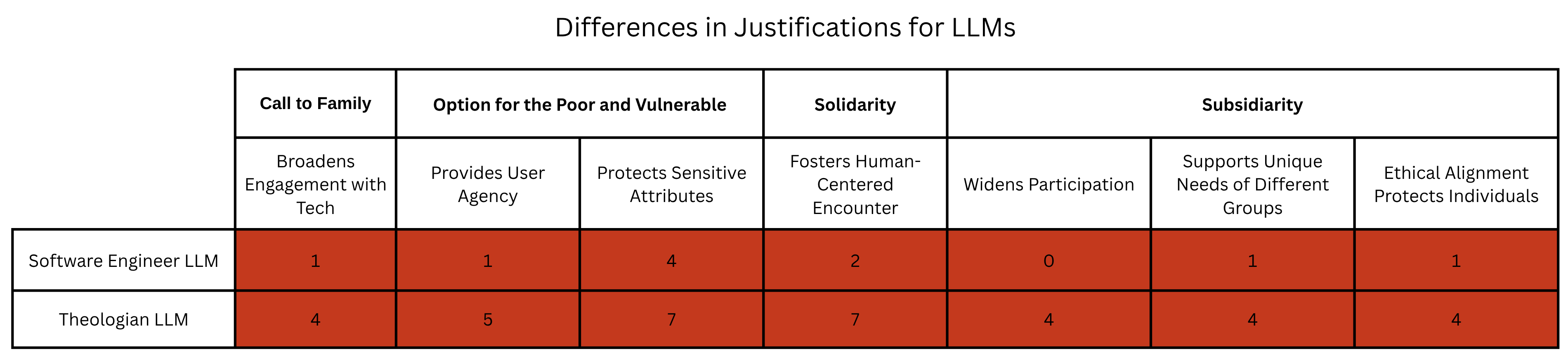}
    \caption{The themes that had the biggest difference between groups in number of participants who mentioned that theme for the LLM design patterns. Any theme that had a between-group difference of 3 or more participants mentioning that theme in their justification is listed here.}
    \label{fig:llm-theme-differences}
    \Description{Table titled "Differences in Justifications for LLMs", comparing responses between Software Engineer LLM and Theologian LLM across four Catholic Social Teaching principles and seven specific sub-themes.
    - Call to Family (Broadens Engagement with Tech): Software Engineer LLM: 1, Theologian LLM: 4.
    - Option for the Poor and Vulnerable (Provides User Agency): Software Engineer LLM: 1, Theologian LLM: 5.
    - Option for the Poor and Vulnerable (Protects Sensitive Attributes): Software Engineer LLM: 4, Theologian LLM: 7.
    - Solidarity (Fosters Human-Centered Encounter): Software Engineer LLM: 2, Theologian LLM: 7.
    - Subsidiarity (Widens Participation): Software Engineer LLM: 0, Theologian LLM: 4.
    - Subsidiarity (Supports Unique Needs of Different Groups): Software Engineer LLM: 1, Theologian LLM: 4.
    - Subsidiarity (Ethical Alignment Protects Individuals): Software Engineer LLM: 1, Theologian LLM: 4.}
\end{figure}

\begin{figure}
    \centering
    \includegraphics[width=\linewidth]{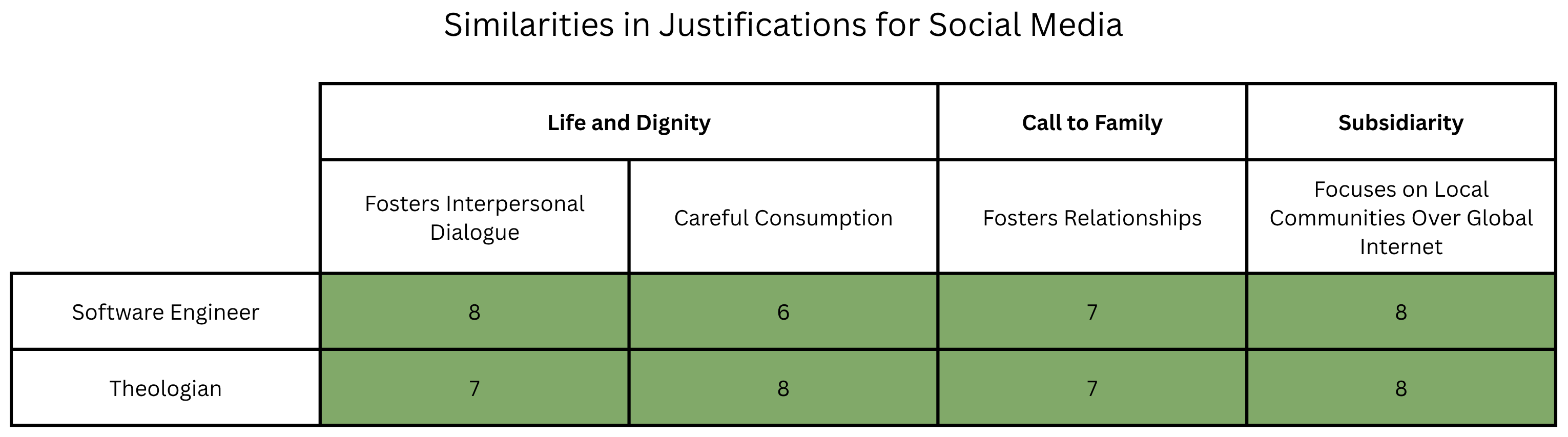}
    \caption{The themes that had the biggest similarity between groups in number of participants who mentioned that theme for the social media design patterns. Any theme that had at least 6 participants in each group is listed here.}
    \label{fig:sm-theme-similarities}
    \Description{Table titled "Similarities in Justifications for Social Media", comparing responses between Software Engineer and Theologian across three Catholic Social Teaching principles and four specific sub-themes.
    - Life and Dignity (Fosters Interpersonal Dialogue): Software Engineer: 8, Theologian: 7.
    - Life and Dignity (Careful Consumption): Software Engineer: 6, Theologian: 8.
    - Call to Family (Fosters Relationships): Software Engineer: 7, Theologian: 7.
    - Subsidiarity (Focuses on Local Communities Over Global Internet): Software Engineer: 8, Theologian: 8.}
\end{figure}

\begin{figure}
    \centering
    \includegraphics[width=\linewidth]{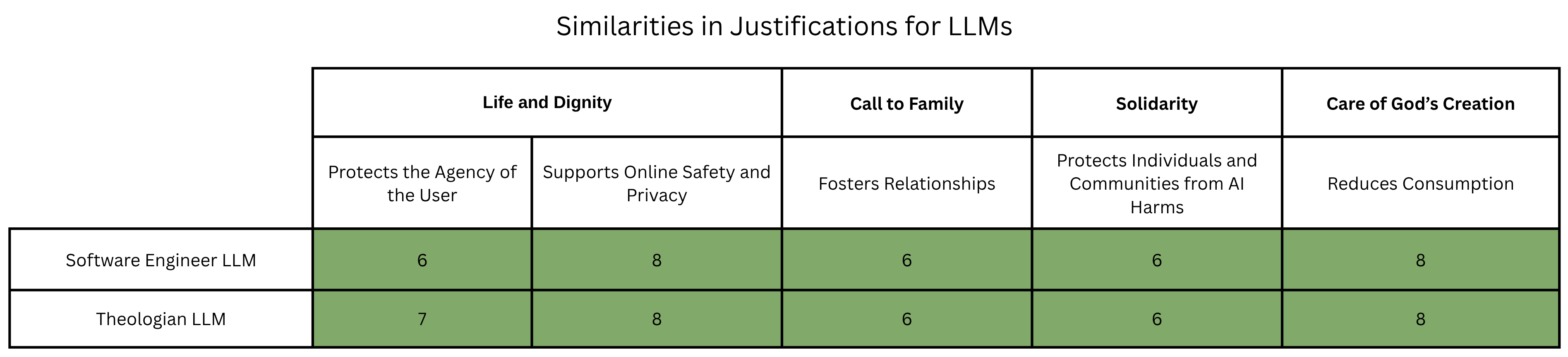}
    \caption{The themes that had the biggest similarity between groups in number of participants who mentioned that theme for the LLM design patterns. Any theme that had at least 6 participants in each group is listed here.}
    \label{fig:llm-theme-similarities}
    \Description{Table titled "Similarities in Justifications for LLMs", comparing responses between Software Engineer LLM and Theologian LLM across four Catholic Social Teaching principles and five specific sub-themes.
    - Life and Dignity (Protects the Agency of the User): Software Engineer LLM: 6, Theologian LLM: 7.
    - Life and Dignity (Supports Online Safety and Privacy): Software Engineer LLM: 8, Theologian LLM: 8.
    - Call to Family (Fosters Relationships): Software Engineer LLM: 6, Theologian LLM: 6.
    - Solidarity (Protects Individuals and Communities from AI Harms): Software Engineer LLM: 6, Theologian LLM: 6.
    - Care of God's Creation (Reduces Consumption): Software Engineer LLM: 8, Theologian LLM: 8.}
\end{figure}

Summary figures of what our research team interpretively assessed to be the areas of biggest difference and biggest similarity between justifications can be seen in figures \ref{fig:sm-theme-differences} - \ref{fig:llm-theme-similarities}. We define ``major difference'' as a between-group difference of 3 or more participants mentioning a particular theme. We define ``major similarity'' as any theme that was mentioned by at least 6 or more participants in both groups.

A figure summarizing the number of times all themes were mentioned by the software engineer and theologian groups can be found in the appendix. Notably there were many instances when the same number of participants per group mentioned the same themes, yet these instances did not fit the inclusion criteria for major similarity because it was fewer than six participants.

We explain the themes that had major similarity or major difference below. The themes that did not fit the criteria for major similarity or major difference are detailed in the appendix.

\subsubsection{Life and Dignity of the Human Person}\mbox{}

\textbf{Protects the Agency of the User} (major \textcolor{red}{\textbf{difference}} for social media, major \textcolor{blue}{\textbf{similarity}} for LLMs). This theme relates to systems that protect user agency through increased transparency. One example is transparent social media recommendation algorithms which T5 connected to human dignity by saying, \textit{``making algorithms customizable...is a really important form of autonomy...it's a big concern that a lot of us have right now about how algorithms are modeling our thinking and behavior in ways that we often have not just little control over, but little awareness of.''} Another element of this theme is LLM chatbots that make it clear they are not humans and avoid sycophancy: \textit{``anthropomorphic AI can deceive users into thinking that they have entered into a relationship with a machine, and can distort their self-understanding and their understanding of reality because of that''} (T7).

\textbf{Fosters Interpersonal Dialogue} (major \textcolor{blue}{\textbf{similarity}} for social media). This theme refers to technologies that promote communication between people, especially in a manner that is respectful and cultivates relationships. This includes social media platforms with features that promote conversation, especially noting that private conversations can lead to increased vulnerability. SE2 explained, \textit{``if you're interacting more with people that you actually know and you actually care about, not some, some influencer...it's not the insecurity based social media that we're so used to.''} This theme can also refer to moderation that ensures respectful conduct through \textit{``moderating away speech and actions that we find to be lacking dignity''}(SE1). Social platforms that limit bots can also promote this through ensuring people are only talking to other humans. In the context of LLMs, this theme encompasses chatbots that don't replace human relationships: \textit{``I think it does humans a disservice when you have...an LLM out there acting as your friend, acting as essentially another human, or a companion...the last thing you want to do is starve humanity of connecting with real people''} (SE4).

\textbf{Careful Consumption} (major \textcolor{blue}{\textbf{similarity}} for social media). This theme refers to technologies that don't encourage excessive consumption of both the technology itself and other goods. This includes social media platforms that don't encourage excessive use through infinite scroll feeds and large numbers of notifications. T6 connected these designs to \textit{life and dignity of the human person} saying, \textit{``In the case of notifications or in the case of algorithms...they are both designed to manipulate me...the manipulation is contrary to the dignity of the human person to treat the person as a thing.''}  This theme also encompasses resisting technology designs that encourage a consumeristic mindset: \textit{``social media [should] be actually social, instead of...an algorithmically curated advertising feed''} (T3). The final dimension of this theme is social media and LLM chatbots that don't use too many environmental resources. About careful context T2 said, \textit{``if you can have the smaller [context window] do exactly what you [need]...it will make less impact on the environment.''}

\textbf{Supports Online Safety and Privacy} (major \textcolor{blue}{\textbf{similarity}} for social media). This theme refers to social media platforms that have moderation to protect users from harmful content or harmful interactions with other people, including \textit{``filtering and making sure conversations stay appropriate and stay centered on the dignity of...people who are very vulnerable''} (T3). This theme also refers to LLM chatbots that are aligned to protect users from harmful responses. This can either mean preventing \textit{``systems that are...inherently trained to be...biased or unfair or stereotypical or racist''} (SE4) or systems that \textit{``make suggestions or be used in such a way that they instruct people on how to harm themselves or harm others''} (T7).

\subsubsection{Call to Family, Community, and Participation}\mbox{}

\textbf{Broadens Engagement with Tech} (major \textcolor{red}{\textbf{difference}} for LLMs). This theme means more people being able to use and customize the technologies. One aspect of this is making LLM chatbots available in more languages: \textit{``different languages can broaden the community that's participating''} (T4). Another element of this is customizing algorithms. This includes making LLM parameters more customizable: T1 said, \textit{``in the sense of the call to participation...[power in the parameters relates] by promoting more of a participatory engagement with the models'' }(T1). Customizing algorithms also includes allowing people to tailor the content in their social media feeds. T5 said about the clear algorithmic comprehension design pattern, \textit{``being able to...exercise some form of control or customization of the algorithms...that's important [to] the notion of participation.''}

\textbf{Fosters Relationships} (major \textcolor{blue}{\textbf{similarity}} for social media and LLMs). This theme refers to social media platforms that encourage active participation in conversations, promote interpersonal communication rather than impersonal scrolling, and chat-based platforms and moderated groups that foster intentional conversations. T3 found feed-based platforms to violate this, saying \textit{``less feed is better...the name feed conjures up feeding livestock...that should tell us that that's really not in keeping with...the way that we want to relate to each other.''} Framed more positively, SE7 reminisced about when chat-based social technologies were more popular: \textit{``we grew up in the time where...we had MSN and...no doom scrolling. Instead of having thousands of followers, you had a few that were actually your good friends that you could actually chat.''} This theme also refers to LLM chatbots that do not replace human relationships. SE2 explained the danger of LLM chatbots replacing human relationships: \textit{``Let's say that if you have a very convincing AI that seems like a person, people might be building relationships with this AI rather than with people they know or with their family, with their friends, with partners.''}

\subsubsection{Option for the Poor and Vulnerable}\mbox{}

\textbf{Provides User Agency} (major \textcolor{red}{\textbf{difference}} for LLMs). This theme refers to greater control over social media algorithms and LLM parameters, increased algorithmic transparency, and both social media and LLM platforms that do not employ manipulative designs. SE1 explained that on social media platforms, \textit{``algorithmic recommendations...are often designed to prey on a fixation of sorts that people have...if people are desperate for money, desperate for help, desperate for connection, they tend to feed that quite a bit in ways that are typically not healthy for people....[clear algorithmic comprehension is] avoiding exploitation...on your platform.''} Other participants highlighted the manipulative nature of humanlike chatbots: \textit{``if we think of...the poor and vulnerable being those who are chronically isolated and depressed, I think the anti-anthropomorphic [design pattern] is important''} (T2).

\textbf{Protects Sensitive Attributes} (major \textcolor{red}{\textbf{difference}} for LLMs). This theme refers to algorithms, especially LLM responses, that limit bias and do not exploit sensitive attributes for predatory targeting. T5 related the clear algorithmic comprehension design pattern to the \textit{option for the poor and vulnerable}, noting \textit{``it's important that there be transparency and it's important that there be critical perspectives that take...subtle forms of discrimination into account.''} T8 noted the importance of the rights reinforcement design pattern, saying \textit{``this idea of...safety, ethical alignment...is particularly important here because it's easy to take advantage of the vulnerable.''}

\subsubsection{Solidarity}\mbox{}

\textbf{Provides Support and Access to Resources} (major \textcolor{red}{\textbf{difference}} for social media). This theme refers to technologies that strengthen community ties, such as social media platforms that foster relationships between people and LLM chatbots that do not detract from building human relationships. This theme also refers to technologies that encourage interactions offline and help connect people to help support each other's needs. T8 gave the following hypothetical example of how online communities might support each other offline: \textit{``if houses burned down in the neighborhood...the neighborhood Facebook group is much more open to helping people out if they know them as, you know, an actual individual, not as, like [anonymous accounts].''}

\textbf{Fosters Human-Centered Encounter} (major \textcolor{red}{\textbf{difference} for LLMs}). This theme refers to social media platforms that promote interpersonal encounter, including by promoting active participation in conversations and by verifying participants to ensure a more safe and intentional community. One example of this is the moderated entry design pattern which SE4 said is not \textit{``to keep people out...moderated entry is being used to build the richest group of people to all experience solidarity together.''} Another example is chat-based social media \textit{``because of the ability to connect....especially if you have different views from...whoever is posting and you want to chat about it''} (T1). This theme also refers to LLM chatbots that don't encourage people to replace human interactions with AI, and LLM translation that helps break language barriers to help foster community. On the anti-anthropomorphic design pattern, T7 said, \textit{``I think treating AI like a person kind of indirectly undermines solidarity between actual human persons…[it] inclines us towards a reductionistic understanding of the human person as just a machine.''}

\textbf{Protects Individuals and Communities from AI Harms} (major \textcolor{blue}{\textbf{similarity}} for LLMs). This theme refers to the fact that AI alignment can protect people from harm, including (but not limited to) high-stakes situations like protecting the environment and using chatbots for wartime decisions. T5 saw the rights reinforcement design pattern as upholding solidarity because \textit{``with the debate about Anthropic...the ethical restrictions that they wanted to put on their product involved the use for automated weapon systems...this kind of question is related to solidarity.''} In a broader sense, the rights reinforcement design pattern upholds solidarity \textit{``in the sense that human rights really are designed to be a kind of tool of solidarity, what we all share in common, what we should never violate''} (T1). Other participants saw the careful context design pattern as upholding solidarity because \textit{``we don't want our AI just destroying our environment''} (SE7).

\subsubsection{Subsidiarity}\mbox{}

\textbf{Supports Unique Needs of Different Groups} (major \textcolor{red}{\textbf{difference}} for social media and LLMs). This theme refers to both online communities having the autonomy to be run the way the members want (especially through self-moderation) and to communities having the ability to customize AI tools and algorithms to fit their needs. Related to moderation T3 said, \textit{``the principle of subsidiarity says we make decisions on the most local level possible and the people who are most greatly impacted by whatever issue we're talking about and trying to decide upon as a group are the ones who have the greatest stake. So having the conversation be guided and led by that group is important.''} Many participants found power in the parameters embodied subsidiarity because it \textit{``would empower smaller communities to leverage the tools better''} (SE7).

\textbf{Widens Participation} (major \textcolor{red}{\textbf{difference}} for LLMs). This theme was only mentioned in the context of the cross-lingual consistency LLM design pattern, and refers to increased language translation ability enabling increased inclusion and societal participation. T3 explained, \textit{``The principle of inclusion and participation is the measure of a just society in Catholic Social Teaching, this sort of undergirds everything that we've been talking about. is facilitated by us being able to talk to each other.''}

\textbf{Ethical Alignment Protects Individuals} (major \textcolor{red}{\textbf{difference}} for LLMs). This theme refers to the top-down dimension of subsidiarity that acknowledges the necessity of higher-levels of control to achieve the common good, and relates this to ethical alignment in software systems. Specifically, this refers to tech companies having the power and responsibility to ethically align AI models, limit harmful AI capabilities, and provide algorithmic transparency to users. T5 noted the rights reinforcement design pattern as an example of this: \textit{``rights reinforcement is a way of ensuring that collective, common good values are preserved...I don't see subsidiarity simply as a value of decentralization. And there are certain conditions where you need to have common, even universal, ethical values upheld for the entire system''} (T5).

\textbf{Focuses on Local Communities Over Global Internet} (major \textcolor{blue}{\textbf{similarity}} for social media). This theme refers to online communities that center around small group, interpersonal interactions, sometimes with members of one's immediate local community. SE5 connected this to subsidiarity saying when \textit{``discourse is happening at the lowest level...the actual power of...that conversation...[is] emergent from lower-level conversation rather than global noise that's amplified.''} These communities are sometimes intentionally bounded to maintain their local character, for example, \textit{``if a particular group of people want to have a conversation about something that is affecting their immediate community, restricting participation to that group could make a lot of sense''} (T3).

\subsubsection{Care of God's Creation}\mbox{} 

\textbf{Reduces Consumption} (major \textcolor{blue}{\textbf{similarity}} for LLMs). This theme refers to technology designs that reduce the consumption of the technology itself and reduce the technology's consumption of environmental resources. SE4 saw \textit{care of God's creation} as \textit{``we're trying to actually be stewards of creation and not use up all of Creation's resources. And along that train of thought, we would want to ensure that the systems we're building are the most efficient and resource, non-intensive as possible.''} It also refers to social media designs that do not promote consumerism, because \textit{``so much of the stuff that we are targeted for [on social media] is to sell stuff that we don't need''} (T1).

\section{Findings Part 2: Reflexivity of Domain}
In this section we report the average number of design patterns selected by theologians and software engineers. We also share observations from the interviews that suggest the number of patterns selected is related to the participant's professional domain. In other words, one's professional experience can impact reflexivity in the the technical inquiry. 

\begin{figure}
    \centering
    \includegraphics[width=\linewidth]{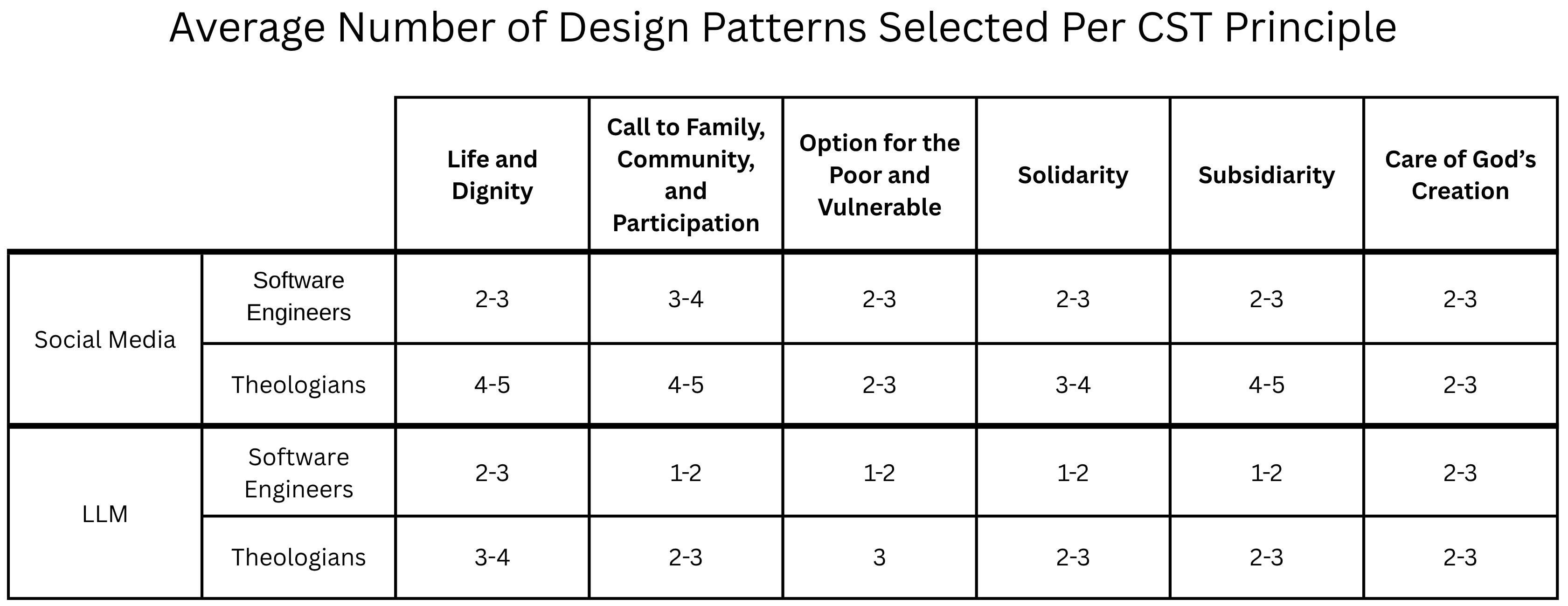}
    \caption{The average number of design patterns selected for a particular CST principle. Decimal averages are converted to a range (e.g. 2.714 is displayed as 2-3). We see that in every case, theologians selected more design patterns than software engineers.}
    \label{fig:average-chosen}
    \Description{Table titled "Average Number of Design Patterns Selected Per CST Principle", showing average selection ranges for Social Media and LLM categories across Software Engineers and Theologians for six principles.
    - Social Media:
    * Software Engineers: Life and Dignity 2–3; Call to Family, Community, and Participation 3–4; Option for the Poor and Vulnerable 2–3; Solidarity 2–3; Subsidiarity 2–3; Care of God's Creation 2–3.
    * Theologians: Life and Dignity 4–5; Call to Family, Community, and Participation 4–5; Option for the Poor and Vulnerable 2–3; Solidarity 3–4; Subsidiarity 4–5; Care of God's Creation 2–3.
    - LLM:
    * Software Engineers: Life and Dignity 2–3; Call to Family, Community, and Participation 1–2; Option for the Poor and Vulnerable 1–2; Solidarity 1–2; Subsidiarity 1–2; Care of God's Creation 2–3.
    * Theologians: Life and Dignity 3–4; Call to Family, Community, and Participation 2–3; Option for the Poor and Vulnerable 3; Solidarity 2–3; Subsidiarity 2–3; Care of God's Creation 2–3.}
\end{figure}

In the design pattern selection task we observed that theologians on average selected more design patterns as embodying a CST principle than software engineers did (see figure \ref{fig:average-chosen}). Theologians on average selected around 2-5 patterns per CST principle whereas software engineers on average selected around 1-3 patterns.

Additional observations from the interviews may shed light on why this occurred. During her interview SE8 made the following offhand comment: ``If I’m developing [according to] some guiding principle...you can only choose one or two [design patterns] otherwise you’re going to do it poorly.'' Rather than selecting every possible design pattern that could relate to the CST principle under consideration, SE8 took other practical constraints of the engineering process into account when making her selections.

Similarly, an unanticipated interview with a historian of CST provided additional context for why the theologians may have selected so many design patterns. For the eighth CST expert interview, the first author initially recruited a historian of CST thinking she was a theologian. During the interview the first author noticed that the historian's responses were outliers in comparison to the previous seven theologian interviews. After the interview completed, the first author discussed this observation with the participant, who said she believed it was because she is a historian and not a theologian. While the theologians selected many design patterns as applying to each CST principle, for almost every CST principle the historian chose none of design patterns as relating. Yet, the interviewer observed in the historian's justifications that she had a similar conceptual understanding of the CST principles that the theologians did, even though her technical selections differed. In a follow-up email correspondence, the historian explained her perceived differences between how she and a theologian would approach our design pattern selection task (reproduced with permission):

\begin{quotation}
    \textit{There are methodological differences between theologians and historians around interpretative practice. As a historian, I’m trained to do my best not to interpret the source I’m provided, but to ground it in its intended context and meaning, and to understand and apply it within that context. Theology - to my limited understanding - has an interpretative, applied element to its practice because it is considered a set of living ideas or beliefs. Theologians, then, (again, from an external limited perspective) are more comfortable seeking or seeing an interpretation or partially applied principle in your examples than I would be - I would be, and did, stick as closely as possible to the CST principles as I understand their original intent in analyzing whether they were applied in practice in your models.}
\end{quotation}

In other words, theologians with an expertise in CST would be more likely to have a broader or more flexible view of what a design pattern that embodies CST could look like, whereas a historian with expertise in CST would be more likely to have a narrower view. Two people with similarly deep understandings of the same values may still have different interpretations because of their professional domain.

\section{Discussion}
We discuss how reflexivity of experience and reflexivity of domain should be considered in design processes, as well as limitations and future work.

\subsection{Designing Considering Reflexivity of Experience: Theologians Demonstrate Broader Understanding than Software Engineers}

Our findings for \textit{reflexivity of experience} demonstrated that although much of the time theologians and software engineers responded similarly, the areas of difference suggest theologians have a broader understanding of CST. While the software engineers primarily relied on the handout to inform their understanding of the CST principles, the theologians also relied on their academic expertise to bring an additional perspective.

Because the majority of the time the groups responded similarly, this suggests that when resources are limited software engineers with little familiarity of a religious value system can fairly effectively design for those values with just a short explanation. In other words, our CST handout was sufficient in many cases to bring the software engineer group quickly up to speed to perform similarly to the theologians. However, integrating theologians into the design process may help to provide the most robust or nuanced understandings of religious principles.

In fact, we can see from many of the major similarities and major differences in justifications how the software engineers relied on the handout: most of the major similarities in justifications relate to content on the handout and most of the major differences relate to interpretations of the CST principles not included on the handout. For example, for \textit{life and dignity of the human person} in the context of social media, the theme \textbf{protects the agency of the user} was a justification with major difference. Agency is not mentioned at all on the CST handout. The theologians frequently mentioned agency in the context of the clear algorithmic comprehension and notification intentionality design patterns, both of which showed significant or possible difference on the Fisher's Exact Test. This shows that this difference in conceptual understanding filtered down to the technical inquiry. Agency was also a justification for \textit{option for the poor and vulnerable} that had major difference between theologians and software engineers, further underscoring how theologians relied on their prior experience with these principles.

Other differences in responses cannot necessarily be explained by a lack of information on the handout. For example, \textbf{protects sensitive attributes} (\textit{option for the poor and vulnerable}), \textbf{provides support and access to resources} (solidarity), and \textbf{fosters human-centered encounter} (\textit{solidarity}) are consistent with the information on the handout. This suggests there may be instances where previous deep engagement with a value system can cultivate a broader moral imagination that helps individuals more easily engage with and apply values in practice.

Because of this, future work should explore the impact of providing software engineers with training in religious values beyond a simple handout, then comparing their outputs to those of theological experts. Classroom education is one approach, but Shilton’s `values levers'~\cite{shilton2013values} suggests that practice-based methods (such as working on interdisciplinary teams) can move beyond traditional approaches centered around classroom education of engineers or bringing value experts onto design teams. Although Shilton focused on motivating engineers to prioritize designing for values, as opposed to our task of teaching engineers about particular values, we can draw inspiration from Shilton's framework.

\subsection{Designing Considering Reflexivity of Domain: Potential Mismatches in the Conceptual and Technical Inquiries}

Our findings suggest that even when conceptual understandings are the same, differences in domain can impact the technical inquiry: in our case, the number of design patterns selected by theologians and software engineers. We also see this in the anecdotal example of the CST historian who in almost every case chose no design patterns or just one design pattern, but had the same academic depth of training in CST as the theologians.

These findings first suggest first that when designing for CST, and perhaps R/S values more broadly, when recruiting value experts it is important to be aware of their domain. There is not a ``right'' or ``wrong'' domain, but rather different domains bring different experiences that could impact the design process. In fact, there could be a benefit in bringing different domains into conversation to dialogue about the best approach for designing for particular values. Second, the fact that reflexivity of domain causes differences primarily in the technical inquiry and not in the conceptual inquiry suggests that attention should be paid at all stages of VSD. For example if a software team decides to consult with religious experts, it may be more valuable to include them in all stages of the design process rather than just bringing them in for the conceptual inquiry to provide value definitions.

\subsection{Limitations and Future Work}
The limitations of this study include its US-dominant perspective, the fact that each participant group only had eight participants, and that we only studied one religious value system. Each of these limitations invites opportunities for future work.

First, although not all of our participants were US citizens, every participant was either a US citizen or a US resident. Software engineers from other cultures or value systems may have had either more similar or more different responses from the CST experts depending on how similar a CST understanding of a value is compared to a similar value in their cultural system. For example, while the principle of subsidiarity had some of the greatest difference in understanding between the software engineer and theologian groups, it's possible that a nonreligious software engineer from Europe may have been more familiar with subsidiarity because it is a part of the European Union's constitution. On the contrary, our software engineers had a very similar understanding of Care of God's Creation to the CST experts, but if we were interviewing a group of software engineers from the Global South their understanding of Care of God's Creation might be different~\cite{preservation2003radical}. Thus there is an important opportunity for further study with non-US populations.

Second, while our reflexive thematic analysis indicated data saturation at eight participants per group, this is a fairly small number of participants. Although we used a Fisher's Exact Test which is designed for use with small populations, our claims of significant difference and possible difference should be interpreted with caution. Future work with more or different participants would be valuable to validate these results or to uncover different findings.

Third, in this study we asked participants to make selections from a set list of design patterns. While this was an intentional choice to create a more controlled environment for comparison between groups, it would also be valuable to host design workshops that allow for more open-ended responses in both the conceptual and technical inquiry stages.

Finally, in this paper we only studied understandings of CST. Because many of the values of CST may be easily translatable to secular American values, there is great value in future work investigating understandings of religious value systems that be less similar or easily translatable to secular values.

\section{Conclusion}
We compared how software engineers with little to no religious affiliation and theologians with academic expertise in CST understood social media and LLM design patterns as embodying different principles of CST. While in many cases the software engineers and theologians responded similarly, the few cases of difference suggested a \textit{reflexivity of experience}: the theologians had a broader understanding of the principles than the software engineers did, possibly stemming from their greater experience with the principles. Our interviews also surfaced a \textit{reflexivity of domain}: one's professional domain impacting the interpretation of values in the technical inquiry. Software teams should thus consider both \textit{reflexivity of experience} and \textit{reflexivity of domain} when designing for R/S values. For those interested in designing for CST specifically, particular care should be taken when designing for \textit{life and dignity of the human person}, \textit{solidarity}, and \textit{subsidiarity}, as these principles had the biggest differences in understanding between the software engineers and the theologian experts. By integrating attentiveness to differences of experience and domain in design processes, software teams will be able to more effectively build systems that authentically embody religious values.

\begin{acks}
\end{acks}

\bibliographystyle{ACM-Reference-Format}
\bibliography{sample-base}

@String{Computing = "Computing" }

@String{Computer = "{IEEE} Computer" }

@String{Springer = "Springer-Verlag" }

@inproceedings{wolf2024still,
author = {Wolf, Sara and Friedrich, Paula and Hurtienne, J\"{o}rn},
title = {Still Not a Lot of Research? Re-Examining HCI Research on Religion and Spirituality},
year = {2024},
isbn = {9798400703317},
publisher = {Association for Computing Machinery},
address = {New York, NY, USA},
url = {https://doi.org/10.1145/3613905.3651058},
doi = {10.1145/3613905.3651058},
booktitle = {Extended Abstracts of the CHI Conference on Human Factors in Computing Systems},
articleno = {302},
numpages = {15},
location = {Honolulu, HI, USA},
series = {CHI EA '24}
}

@inproceedings{kitson2025diving,
author = {Kitson, Alexandra and Halperin, Brett A. and Smith, C. Estelle and Maas, Franzisca and Hoefer, Michael J and Wolf, Sara and Buie, Elizabeth},
title = {Diving Into the Ocean of Religious and Spiritual Design Research},
year = {2025},
isbn = {9798400714863},
publisher = {Association for Computing Machinery},
address = {New York, NY, USA},
url = {https://doi.org/10.1145/3715668.3734166},
doi = {10.1145/3715668.3734166},
booktitle = {Companion Publication of the 2025 ACM Designing Interactive Systems Conference},
pages = {47–51},
numpages = {5},
location = {
},
series = {DIS '25 Companion}
}

@inproceedings{wolf2026nospirituality,
author = {Wolf, Sara and Friedrich, Paula and Buie, Elizabeth and Blythe, Mark},
title = {No Spirituality Please, We’re HCI: Challenges for HCI Research on Religion and Spirituality},
year = {2026},
isbn = {9798400722783},
publisher = {Association for Computing Machinery},
address = {New York, NY, USA},
url = {https://doi.org/10.1145/3772318.3790490},
doi = {10.1145/3772318.3790490},
booktitle = {Proceedings of the 2026 CHI Conference on Human Factors in Computing Systems},
articleno = {1265},
numpages = {29},
location = {Barcelona, Spain},
series = {CHI '26}
}

@inproceedings{buie2013spirituality,
author = {Buie, Elizabeth and Blythe, Mark},
title = {Spirituality: there's an app for that! (but not a lot of research)},
year = {2013},
isbn = {9781450319522},
publisher = {Association for Computing Machinery},
address = {New York, NY, USA},
url = {https://doi.org/10.1145/2468356.2468754},
doi = {10.1145/2468356.2468754},
booktitle = {CHI '13 Extended Abstracts on Human Factors in Computing Systems},
pages = {2315–2324},
numpages = {10},
location = {Paris, France},
series = {CHI EA '13}
}

@misc{germain2023,
    title={The Vatican Releases Its Own AI Ethics Handbook},
    url={https://gizmodo.com/pope-francis-vatican-releases-ai-ethics-1850583076},
    journal={Gizmodo},
    author={Germain, Thomas},
    year={2023},
    month={Jun}
}

@misc{romecall,
    title={Rome Call for AI Ethics},
    url={https://www.romecall.org},
    year={2020},
    month={Feb},
    author={RenAIssance Foundation}
}

@misc{anthropicmagnifica,
    title={Anthropic co-founder Chris Olah's remarks on Pope Leo XIV's encyclical "Magnifica humanitas"},
    url={https://www.anthropic.com/news/chris-olah-pope-leo-encyclical},
    year={2026},
    month={May},
    author={Anthropic}
}

@article{tiku2026can,
  author       = {Tiku, Nitasha},
  title        = {Can {AI} be a `child of {G}od'? {I}nside {A}nthropic's meeting with {C}hristian leaders},
  journal      = {The Washington Post},
  year         = {2026},
  month        = {April},
  day          = {11},
  url          = {https://www.washingtonpost.com/technology/2026/04/11/anthropic-christians-claude-morals/},
  note         = {Accessed: July 14, 2026}
}

@misc{fauria2026openai,
  author       = {Fauria, Krysta},
  title        = {OpenAI and Anthropic just met with religious leaders at the 'Faith-AI Covenant.' Here's why},
  howpublished = {Fast Company},
  month        = {may},
  year         = {2026},
  url          = {https://www.fastcompany.com/91538977/openai-anthropic-just-met-religious-leaders-faith-ai-covenant-heres-why},
  note         = {Accessed: July 14, 2026}
}

@article{timmermans2014reflexivity,
  title={Reflexivity and value-sensitive design},
  author={Timmermans, Job and Mittelstadt, Brent},
  journal={CEPE 2014 Proceedings},
  year={2014}
}

@article{griffin2025ethical,
  title={The ethical wisdom of AI developers},
  author={Griffin, Tricia A and Green, Brian P and Welie, Jos VM},
  journal={AI and Ethics},
  volume={5},
  number={2},
  pages={1087--1097},
  year={2025},
  publisher={Springer}
}

@inproceedings{conwill2025design,
author = {Conwill, Louisa and Levis, Megan K. and Badillo-Urquiola, Karla and Scheirer, Walter J.},
title = {Design Patterns for the Common Good: Building Better Technologies Using the Wisdom of Virtue Ethics},
year = {2025},
isbn = {9798400713941},
publisher = {Association for Computing Machinery},
address = {New York, NY, USA},
url = {https://doi.org/10.1145/3706598.3713546},
doi = {10.1145/3706598.3713546},
booktitle = {Proceedings of the 2025 CHI Conference on Human Factors in Computing Systems},
articleno = {831},
numpages = {23},
location = {
},
series = {CHI '25}
}

@article{lad2026possible,
  title={Is It Possible to Make Chatbots Virtuous? Investigating a Virtue-Based Design Methodology Applied to LLMs},
  author={Lad, Matthew P and Conwill, Louisa and Scheirer, Megan Levis},
  journal={arXiv preprint arXiv:2602.03155},
  year={2026}
}

@book{conwill2024virtue,
    address={Collegeville, Minnesota},
    title={Virtue in Virtual Spaces: Catholic Social Teaching and Technology},
    publisher={Liturgical Press},
    author={Conwill, Louisa and Levis, Megan and Scheirer, Walter J.},
    year={2024}
}

@article{braun2019reflecting,
  title={Reflecting on reflexive thematic analysis},
  author={Braun, Virginia and Clarke, Victoria},
  journal={Qualitative research in sport, exercise and health},
  volume={11},
  number={4},
  pages={589--597},
  year={2019},
  publisher={Taylor \& Francis}
}

@book{friedman2019value,
    address={Cambridge, Massachusetts},
    title={Value Sensitive Design: Shaping Technology With Moral Imagination},
    publisher={MIT Press},
    author={Friedman, Batya and Hendry, David},
    year={2019}
}

@inbook{friedman2013value,
    address={Dordrecht},
    title={Value Sensitive Design and Information Systems},
    ISBN={978-94-007-7844-3},
    url={https://doi.org/10.1007/978-94-007-7844-3_4},
    DOI={10.1007/978-94-007-7844-3_4},
    booktitle={Early engagement and new technologies: Opening up the laboratory},
    publisher={Springer Netherlands},
    author={Friedman, Batya and Kahn, Peter H. and Borning, Alan and Huldtgren, Alina},
    year={2013},
    pages={55–95}
}

@inproceedings{alsheikh2011whose,
    author = {Alsheikh, Tamara and Rode, Jennifer A. and Lindley, Si\^{a}n E.},
    title = {(Whose) value-sensitive design: a study of long- distance relationships in an Arabic cultural context},
    year = {2011},
    isbn = {9781450305563},
    publisher = {Association for Computing Machinery},
    address = {New York, NY, USA},
    url = {https://doi.org/10.1145/1958824.1958836},
    doi = {10.1145/1958824.1958836},
    booktitle = {Proceedings of the ACM 2011 Conference on Computer Supported Cooperative Work},
    pages = {75–84},
    numpages = {10},
    location = {Hangzhou, China},
    series = {CSCW '11}
}

@inproceedings{borning2012next,
  title={Next steps for value sensitive design},
  author={Borning, Alan and Muller, Michael},
  booktitle={Proceedings of the SIGCHI conference on human factors in computing systems},
  pages={1125--1134},
  year={2012}
}

@inproceedings{ahman_Thoren_2025,
    address={Cham},
    title={Grounding Existential HCI: A Structured Literature Review on the Study of Technology, Religion, and Spirituality},
    ISBN={978-3-031-93864-1},
    booktitle={Human-Computer Interaction},
    publisher={Springer Nature Switzerland},
    author={Åhman, Henrik and Thorén, Claes},
    editor={Kurosu, Masaaki and Hashizume, Ayako},
    year={2025},
    pages={189–210}
}

@inproceedings{rifat2022integrating,
  title={Integrating religion, faith, and spirituality in HCI},
  author={Rifat, Mohammad Rashidujjaman and Peer, Firaz Ahmed and Rabaan, Hawra and Mim, Nusrat Jahan and Mustafa, Maryam and Toyama, Kentaro and Markum, Robert B and Buie, Elizabeth and Hammer, Jessica and Sultana, Sharifa and others},
  booktitle={CHI conference on human factors in computing systems extended abstracts},
  pages={1--6},
  year={2022}
}

@inproceedings{smith2026spirit,
    author = {Smith, C. Estelle and Bezabih, Alemitu and Nourriz, Shadi and Ovi, Jesan Ahammed},
    title = {SPIRIT: A Design Framework To Support Technology Interventions for Spiritual Care Within and Beyond the Clinic},
    year = {2026},
    isbn = {9798400722783},
    publisher = {Association for Computing Machinery},
    address = {New York, NY, USA},
    url = {https://doi.org/10.1145/3772318.3790662},
    doi = {10.1145/3772318.3790662},
    booktitle = {Proceedings of the 2026 CHI Conference on Human Factors in Computing Systems},
    articleno = {1266},
    numpages = {18},
    location = {Barcelona, Spain},
    series = {CHI '26}
}

@article{ahmed2022situating,
  title={Situating ethics: A postsecular perspective for HCI},
  author={Ahmed, Syed Ishtiaque},
  journal={Interactions},
  volume={29},
  number={4},
  pages={84--86},
  year={2022},
  publisher={ACM New York, NY, USA}
}

@article{goodman1961snowball,
  title={Snowball sampling},
  author={Goodman, Leo A},
  journal={The annals of mathematical statistics},
  pages={148--170},
  year={1961},
  publisher={JSTOR}
}

@article{bezabih2025meeting,
    author = {Bezabih, Alemitu and Nourriz, Shadi and Snider, Anne-Marie and Rauenzahn, Rosalie and Handzo, George and Smith, C. Estelle},
    title = {Meeting Patients Where They're At: Toward the Expansion of Chaplaincy Care into Online Spiritual Care Communities},
    year = {2025},
    issue_date = {November 2025},
    publisher = {Association for Computing Machinery},
    address = {New York, NY, USA},
    volume = {9},
    number = {7},
    url = {https://doi.org/10.1145/3757492},
    doi = {10.1145/3757492},
    journal = {Proc. ACM Hum.-Comput. Interact.},
    month = oct,
    articleno = {CSCW311},
    numpages = {38}
}

@inproceedings{wester2026chaplains,
    author = {Wester, Joel and Rhys Cox, Samuel and Pohl, Henning and van Berkel, Niels},
    title = {Chaplains' Reflections on the Design and Usage of AI for Conversational Care},
    year = {2026},
    isbn = {9798400722783},
    publisher = {Association for Computing Machinery},
    address = {New York, NY, USA},
    url = {https://doi.org/10.1145/3772318.3790468},
    doi = {10.1145/3772318.3790468},
    booktitle = {Proceedings of the 2026 CHI Conference on Human Factors in Computing Systems},
    articleno = {909},
    numpages = {15},
    location = {Barcelona, Spain},
    series = {CHI '26}
}

@article{saunders2018saturation,
  title={Saturation in qualitative research: exploring its conceptualization and operationalization},
  author={Saunders, Benjamin and Sim, Julius and Kingstone, Tom and Baker, Shula and Waterfield, Jackie and Bartlam, Bernadette and Burroughs, Heather and Jinks, Clare},
  journal={Quality \& quantity},
  volume={52},
  pages={1893--1907},
  year={2018},
  publisher={Springer}
}

@inproceedings{lutz2026redteaming,
    author = {Lutz, Nina and Weyl, E. Glen},
    title = {Red teaming with faith leaders: expanding digital safety and accountability to frontiers of care},
    year = {2026},
    isbn = {9798400725968},
    publisher = {Association for Computing Machinery},
    address = {New York, NY, USA},
    url = {https://doi.org/10.1145/3805689.3812362},
    doi = {10.1145/3805689.3812362},
    booktitle = {Proceedings of the 2026 ACM Conference on Fairness, Accountability, and Transparency},
    pages = {399–422},
    numpages = {24},
    location = {
    },
    series = {FAccT '26}
}

@article{lutz2025goodfaith,
  title={Good Faith Design: Religion as a Resource for Technologists},
  author={Lutz, Nina and Olsen, Benjamin and Liu, Weishung and Weyl, E Glen},
  journal={arXiv preprint arXiv:2511.05819},
  year={2025}
}

@article{hiniker2022reclaiming,
    author = {Hiniker, Alexis and Wobbrock, Jacob O.},
    title = {Reclaiming attention: Christianity and HCI},
    year = {2022},
    issue_date = {July - August 2022},
    publisher = {Association for Computing Machinery},
    address = {New York, NY, USA},
    volume = {29},
    number = {4},
    issn = {1072-5520},
    url = {https://doi.org/10.1145/3538706},
    doi = {10.1145/3538706},
    journal = {Interactions},
    month = jun,
    pages = {40–44},
    numpages = {5}
}

@article{hammer2022individual,
    author = {Hammer, Jessica and Reig, Samantha},
    title = {From individual rights to community obligations: a Jewish approach to speech},
    year = {2022},
    issue_date = {July - August 2022},
    publisher = {Association for Computing Machinery},
    address = {New York, NY, USA},
    volume = {29},
    number = {4},
    issn = {1072-5520},
    url = {https://doi.org/10.1145/3535271},
    doi = {10.1145/3535271},
    journal = {Interactions},
    month = jun,
    pages = {30–34},
    numpages = {5}
}

@online{aiandfaith,
  author  = {{AI and Faith}},
  title   = {{AI and Faith}: Religion and the Ethics of {AI}},
  url     = {https://aiandfaith.org/},
  urldate = {2026-08-11}
}

@online{fftn,
  author  = {{Faith Family Technology Network}},
  title   = {{Faith Family Technology Network}},
  url     = {https://www.faithfamilytech.org/},
  urldate = {2026-08-11}
}

@online{buddhismAndAI,
  author  = {{Buddhism \& AI Initiative}},
  title   = {{Buddhism \& AI Initiative}},
  url     = {https://www.engagedbuddhists.ai/},
  urldate = {2026-08-11}
}

@online{spiritedhci,
  title        = {{SPIRITED Collective}: A Research Community at the Intersection of Spirituality, Religion, and Design},
  author       = {{SPIRITED Collective}},
  year         = {2026},
  url          = {https://spiritedhci.org/},
  urldate      = {2026-08-11}
}

@inproceedings{markum2024navigating,
    author = {Markum, Robert B. and Maas, Franzisca and Wolf, Sara and Halperin, Brett A. and Claisse, Caroline and Buie, Elizabeth},
    title = {Navigating Intersections of Religion/Spirituality and Human-Computer Interaction},
    year = {2024},
    isbn = {9798400709654},
    publisher = {Association for Computing Machinery},
    address = {New York, NY, USA},
    url = {https://doi.org/10.1145/3677045.3685453},
    doi = {10.1145/3677045.3685453},
    booktitle = {Adjunct Proceedings of the 2024 Nordic Conference on Human-Computer Interaction},
    articleno = {38},
    numpages = {4},
    location = {Uppsala, Sweden},
    series = {NordiCHI '24 Adjunct}
}

@inproceedings{markum2023designing,
    author = {Markum, Robert B. and Wolf, Sara and Hoefer, Michael and Maas, Franzisca},
    title = {Designing Tangible Interactive Artifacts for Religious and Spiritual Purposes},
    year = {2023},
    isbn = {9781450398985},
    publisher = {Association for Computing Machinery},
    address = {New York, NY, USA},
    url = {https://doi.org/10.1145/3563703.3591463},
    doi = {10.1145/3563703.3591463},
    booktitle = {Companion Publication of the 2023 ACM Designing Interactive Systems Conference},
    pages = {117–120},
    numpages = {4},
    location = {Pittsburgh, PA, USA},
    series = {DIS '23 Companion}
}

@inproceedings{smith2024undesigning,
    author = {Smith, C. Estelle and Bezabih, Alemitu and Freed, Diana and Halperin, Brett A. and Wolf, Sara and Claisse, Caroline and Li, Jingjin and Hoefer, Michael and Rifat, Mohammad Rashidujjaman},
    title = {(Un)designing AI for Mental and Spiritual Wellbeing},
    year = {2024},
    isbn = {9798400711145},
    publisher = {Association for Computing Machinery},
    address = {New York, NY, USA},
    url = {https://doi.org/10.1145/3678884.3689132},
    doi = {10.1145/3678884.3689132},
    booktitle = {Companion Publication of the 2024 Conference on Computer-Supported Cooperative Work and Social Computing},
    pages = {117–120},
    numpages = {4},
    location = {San Jose, Costa Rica},
    series = {CSCW Companion '24}
}

@inproceedings{noor2026decoding,
    author = {Noor, Jannatun and Biswas, Susmita},
    title = {Decoding Islamic HCI: What Current Patterns Reveal About Future Possibilities},
    year = {2026},
    isbn = {9798400722783},
    publisher = {Association for Computing Machinery},
    address = {New York, NY, USA},
    url = {https://doi.org/10.1145/3772318.3791954},
    doi = {10.1145/3772318.3791954},
    booktitle = {Proceedings of the 2026 CHI Conference on Human Factors in Computing Systems},
    articleno = {1582},
    numpages = {23},
    location = {
    },
    series = {CHI '26}
}

@inproceedings{campbell2026codesigning,
    author = {Campbell-Esen, Rochelle and Mahoney, Jamie and Talhouk, Reem},
    title = {Co-Designing Islamic AI Ethics: Insights from the UK Muslim Community},
    year = {2026},
    isbn = {9798400722783},
    publisher = {Association for Computing Machinery},
    address = {New York, NY, USA},
    url = {https://doi.org/10.1145/3772318.3790413},
    doi = {10.1145/3772318.3790413},
    booktitle = {Proceedings of the 2026 CHI Conference on Human Factors in Computing Systems},
    articleno = {930},
    numpages = {18},
    location = {
    },
    series = {CHI '26}
}

@inproceedings{jahan2026progress,
    author = {Jahan, Ishrat and Noor, Jannatun and Nurain, Novia and Islam, A. B. M. Alim Al},
    title = {"It should help me progress and strengthen my Imaan": Understanding Algorithmic Experiences of Islamic Content on Social Media Recommendation Systems},
    year = {2026},
    isbn = {9798400722783},
    publisher = {Association for Computing Machinery},
    address = {New York, NY, USA},
    url = {https://doi.org/10.1145/3772318.3791858},
    doi = {10.1145/3772318.3791858},
    booktitle = {Proceedings of the 2026 CHI Conference on Human Factors in Computing Systems},
    articleno = {601},
    numpages = {18},
    location = {
    },
    series = {CHI '26}
}

@inproceedings{smith2026blackchristian,
    author = {Smith, Alexa N. and Otoo, Nissi and Beach, Rohin and Eaves, Jessie and Jennings, Kaiya and Lyons, George W. and Thompson, Dean and McCall, Terika and O'Leary, Teresa K. and Finda Ogbonnaya-Ogburu, Ihudiya},
    title = {Understanding Digital Religion in the Lives of Black Christian Young Adults},
    year = {2026},
    isbn = {9798400722783},
    publisher = {Association for Computing Machinery},
    address = {New York, NY, USA},
    url = {https://doi.org/10.1145/3772318.3790509},
    doi = {10.1145/3772318.3790509},
    booktitle = {Proceedings of the 2026 CHI Conference on Human Factors in Computing Systems},
    articleno = {1268},
    numpages = {18},
    location = {
    },
    series = {CHI '26}
}

@misc{wingate2026omissive,
      title={Omissive Bias in Religious Representation: Benchmarking LLM Answers to Everyday Ethical Decision-making}, 
      author={David Wingate and Sheryl Carty and Joshua Coates and Daniel Feldman and Nancy Fulda and Larry Howell and Brett Israelson and Dallin Jacobs and Jonathan Karr and John Paul Kimes and Elisabeth Kincaid and Paul Martens and Gavin Mobley and Suzana Pinheiro and Lindsay Slemboski and Peter Whiting},
      year={2026},
      eprint={2605.24319},
      archivePrefix={arXiv},
      primaryClass={cs.LG},
      url={https://arxiv.org/abs/2605.24319}, 
}

@misc{kadous2026jaleesbench,
      title={JaleesBench: Are AI Assistants Good Spiritual Company?}, 
      author={M. Waleed Kadous and Benjamin Olsen},
      year={2026},
      eprint={2608.07508},
      archivePrefix={arXiv},
      primaryClass={cs.HC},
      url={https://arxiv.org/abs/2608.07508}, 
}

@misc{magnificaHumanitas,
  author       = {{Pope Leo XIV}},
  title        = {Encyclical Letter \textit{Magnifica Humanitas}: On Safeguarding the Human Person in the Time of Artificial Intelligence},
  year         = {2026},
  month        = may,
  day          = {15},
  publisher    = {Libreria Editrice Vaticana},
  url          = {https://www.vatican.va/content/leo-xiv/en/encyclicals/documents/20260515-magnifica-humanitas.html},
  note         = {Accessed: 2026-09-01}
}

@book{gamma1995pattern,
    address={Boston},
    title={Design Patterns: Elements of Reusable Object-Oriented Software},
    publisher={Addison-Wesley},
    author={Gamma, Erich and Helm, Richard and Johnson, Ralph and Vlissides, John},
    year={1994}
}

@online{ems2026amish,
  author       = {Lindsay Ems},
  title        = {How the Amish Let Technology in Without Letting It Take Over},
  journaltitle = {The MIT Press Reader},
  date         = {2026-07-27},
  year         = {2026},
  month        = {jul},
  day          = {27},
  url          = {https://thereader.mitpress.mit.edu/how-the-amish-let-technology-in-without-letting-it-take-over/},
  urldate      = {2026-09-01}
}

@online{myjewishlearning_electricity_shabbat,
  author       = {{My Jewish Learning}},
  title        = {Electricity on Shabbat},
  website      = {My Jewish Learning},
  url          = {https://www.myjewishlearning.com/article/electricity-on-shabbat/},
  urldate      = {2026-09-01}
}

@book{compendium,
  author = {Pontifical Council for Justice and Peace},
  title = {Compendium of the Social Doctrine of the Church},
  publisher = {USCCB Publishing},
  year = {2004},
}

@online{gassho2026socialmedia,
  author       = {{Gassho Editorial Team}},
  title        = {How to Use Social Media Mindfully},
  year         = {2026},
  month        = mar,
  day          = {25},
  url          = {https://gassho.info/blog-page/social-media-buddhism/},
  organization = {GASSHO},
  urldate      = {2026-09-01}
}

@article{upton1992fisher,
  title={Fisher's exact test},
  author={Upton, Graham JG},
  journal={Journal of the Royal Statistical Society: Series A (Statistics in Society)},
  volume={155},
  number={3},
  pages={395--402},
  year={1992},
  publisher={Wiley Online Library}
}

@book{hosmer2013applied,
  author    = {Hosmer, David W. and Lemeshow, Stanley and Sturdivant, Rodney X.},
  title     = {Applied Logistic Regression},
  edition   = {3rd},
  publisher = {John Wiley \& Sons},
  year      = {2013},
  doi       = {10.1002/9781118548387}
}

@article{shilton2013values,
  title={Values levers: Building ethics into design},
  author={Shilton, Katie},
  journal={Science, Technology, \& Human Values},
  volume={38},
  number={3},
  pages={374--397},
  year={2013},
  publisher={Sage Publications Sage CA: Los Angeles, CA}
}

@article{preservation2003radical,
  title={RADICAL AMERICAN ENVIRONMENTALISM AND},
  author={PRESERVATION, WILDERNESS},
  journal={Environmentalism: Critical Concepts},
  volume={2},
  number={1},
  pages={58},
  year={2003},
  publisher={Taylor \& Francis}
}

@online{fauria2026pope,
  author       = {Fauria, Krysta},
  title        = {Pope Leo's AI manifesto sparks viral reactions on social media},
  howpublished = {\url{https://apnews.com/article/pope-leo-ai-encyclical-reaction-1abe34ace4705d0c005da4ff85624afa}},
  organization = {Associated Press},
  year         = {2026},
  url          = {https://apnews.com/article/pope-leo-ai-encyclical-reaction-1abe34ace4705d0c005da4ff85624afa},
  urldate      = {2026-09-10}
}

\appendix

\section{Full Interview Protocol}

\textbf{Introduction}\\
Thank you so much for your willingness to be interviewed today. In light of recent efforts using religious principles to inform technology ethics, for example Anthropic consulting Christian leaders, the goal of this interview is to learn how \textbf{[computer scientists with no religious affiliation] / [theologians with no background in computer science]} think about incorporating religious values into design.\\

\noindent Before we begin, I want to make sure you had a chance to review the consent form I sent earlier. Did you get a chance to read it and do you have any questions?\\
\textbf{[Response]}\\

\noindent Great! As the form mentions I would like to take an audio recording of this interview so I can better remember your responses. Do I have your consent to record?\\
\textbf{[Response]}\\

\noindent Thank you. You are free to stop the interview at any point.\\

\noindent \textbf{Background}\\
Before we begin the interview, I’d like to collect some optional demographic information. If you are comfortable sharing, could you please share your age, gender, race/ethnicity, and current role in your job/how long you’ve been in the industry?\\

\noindent \textbf{Initial Questions}\\
\noindent What do you think about technology companies incorporating religious principles into their understanding of ethics?\\

\noindent \textbf{Design Patterns Experiment}\\
\noindent Thank you. To start, I’m going to show you 7 design patterns for social media. If you’re not familiar with the term design pattern, what you need to know for this exercise is that it’s like a written description of a technology design. These design patterns were created to embody particular principles of Catholic Social Teaching, which is the Catholic Church’s teaching on human dignity and societal good, and has a rich commentary on technology’s impact on society.\\

\noindent We want to understand how \textbf{[computer scientists with no experience with CST] / [CST experts with no experience with software design]} see these design patterns as embodying particular principles of CST. I’ll first show you the seven design patterns, and we’ll talk through each one of them to make sure you have an understanding of what it’s doing. Then, I’ll go through six of the main principles of CST and ask you if you wanted to design for that principle, which design pattern or patterns would you use and why. I’ll give you a handout with explanations of each of the principles of CST \textbf{[for theologians: but you are welcome to rely on your previous understanding of these principles as well].} Please now take a few minutes to read through the seven social media design patterns and let me know if you have any questions.\\
\noindent \textbf{[Response…interviewer responds if necessary]}\\

\noindent \textbf{[For each of the six principles…]}

\noindent We’ll consider the principle of \textbf{[insert CST principle].} Take a moment to read through the definition on the handout. Do you have any questions about the definition of the principle?\\
\textbf{[Response…interviewer responds if necessary]}\\

\noindent Which design pattern or patterns would you choose if you wanted to design for this principle? \\
\noindent \textbf{[Response]}\\

\noindent \textbf{[For each pattern selected]} \\
\noindent Why would you choose this pattern to embody \textbf{[insert CST principle]?}\\

\noindent Now, we will do the same exercise but with 5 design patterns for large language model-enabled chatbots. \\

\noindent \textbf{[Same protocol repeats]}

\section{Catholic Social Teaching Infographic}

\begin{figure}
    \centering
    \includegraphics[width=\linewidth]{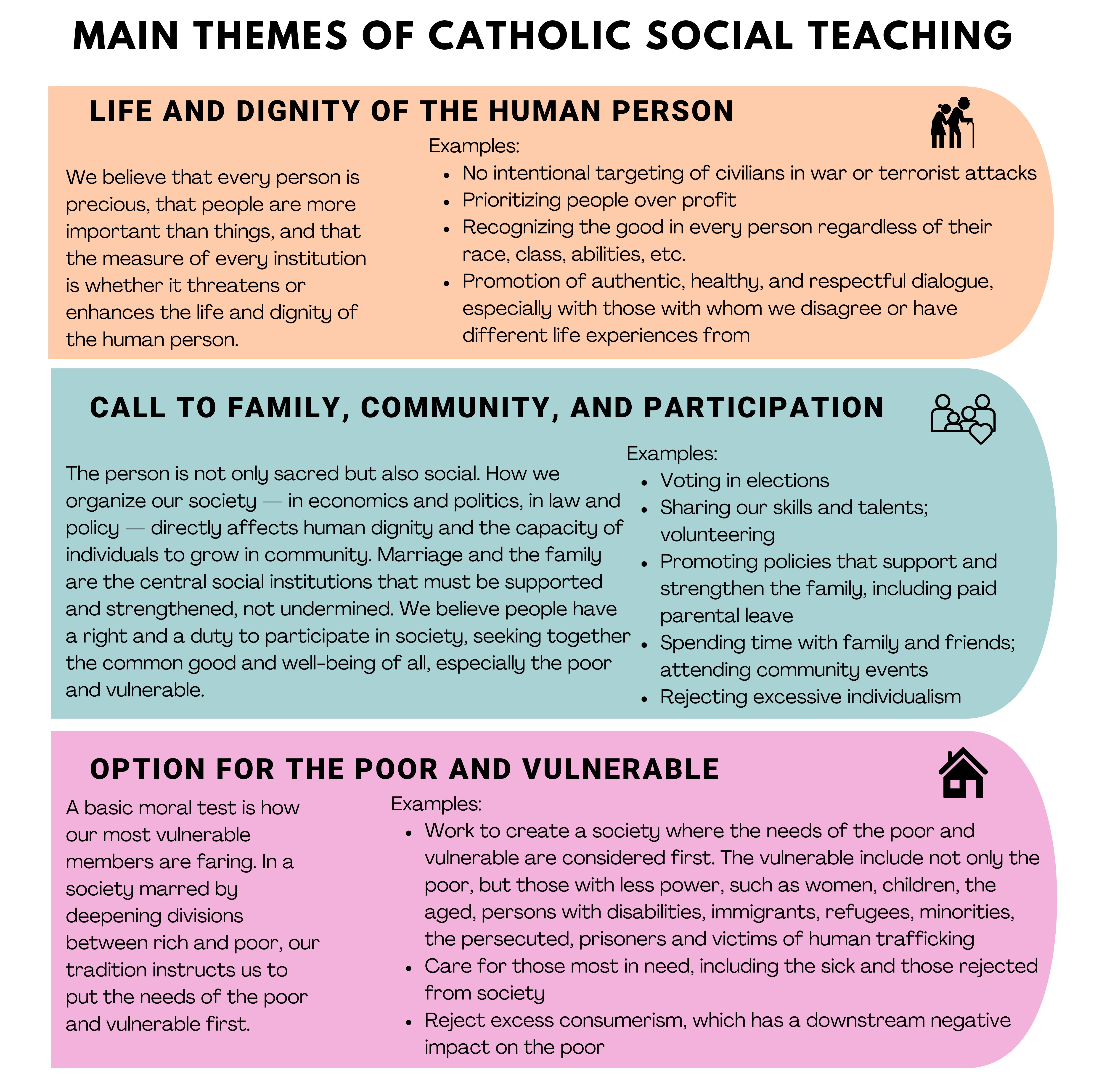}
    \caption{The first page of the Catholic Social Teaching reference handout we gave to the interview participants.}
    \label{fig:infographic-p-1}
    \Description{The text in this figure says, ``Life and dignity of the human person. We believe that every person is precious, that people are more important than things, and that the measure of every institution is whether it threatens or enhances the life and dignity of the human person. Examples: No intentional targeting of civilians in war or terrorist attacks; Prioritizing people over profit; Recognizing the good in every person regardless of their race, class, abilities, etc.; Promotion of authentic, healthy, and respectful dialogue, especially with those with whom we disagree or have different life experiences from.'' There is an icon of two cartoon people hunched over, one with a cane, indicating they are elderly. Next it says, ``Call to family, community, and participation. The person is not only sacred but also social. How we organize our society — in economics and politics, in law and policy — directly affects human dignity and the capacity of individuals to grow in community. Marriage and the family are the central social institutions that must be supported and strengthened, not undermined. We believe people have a right and a duty to participate in society, seeking together the common good and well-being of all, especially the poor and vulnerable. Examples: voting in elections, sharing our skills and talents, volunteering, promoting policies that support and strengthen the family, including paid parental leave, spending time with family and friends, attending community events, rejecting excess individualism.'' There is an icon of two stick figure adults and two stick figure children with a heart, indicating a family. Next it says, ``option for the poor and vulnerable. A basic moral test is how our most vulnerable members are faring. In a society marred by deepening divisions between rich and poor, our tradition instructs us to put the needs of the poor and vulnerable first. Examples: Examples: Work to create a society where the needs of the poor and vulnerable are considered first. The vulnerable include not only the poor, but those with less power, such as women, children, the aged, persons with disabilities, immigrants, refugees, minorities, the persecuted, prisoners and victims of human trafficking. Care for those most in need, including the sick and those rejected from society. Reject excess consumerism, which has a downstream negative impact on the poor.'' There is an icon of a house.
}
\end{figure}

\begin{figure}
    \centering
    \includegraphics[width=\linewidth]{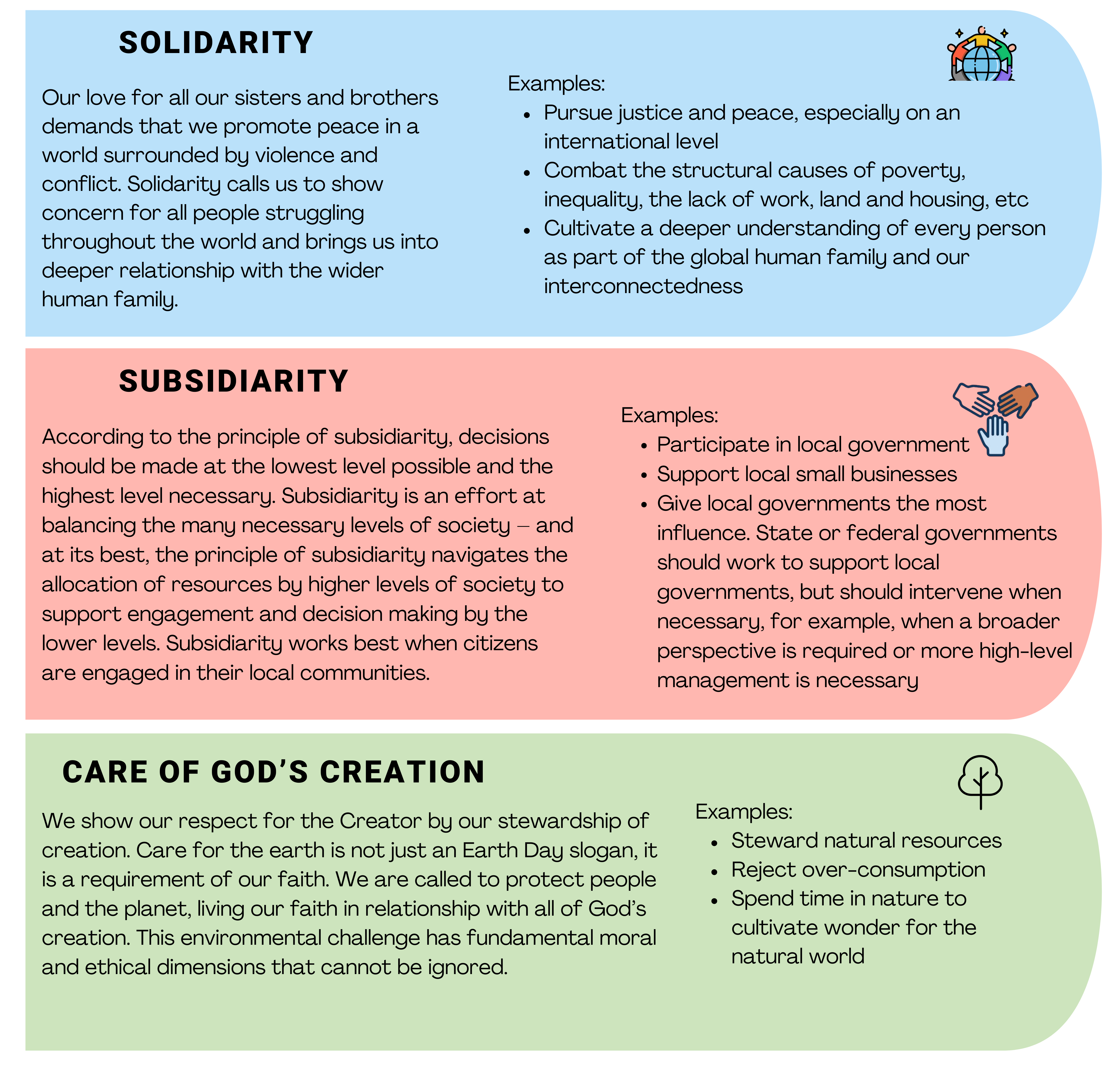}
    \caption{The second page of the Catholic Social Teaching reference handout we gave to the interview participants.}
    \label{fig:infographic-p-2}
    \Description{The text in the figure says, ``Solidarity. Our love for all our sisters and brothers demands that we promote peace in a world surrounded by violence and conflict. Solidarity calls us to show concern for all people struggling throughout the world and brings us into deeper relationship with the wider human family. Examples: Pursue justice and peace, especially on an international level. Combat the structural causes of poverty, inequality, the lack of work, land and housing, etc. Cultivate a deeper understanding of every person as part of the global human family and our interconnectedness.'' There is an icon of a globe with cartoon people holding hands around it. Next it says, ``Subsidiarity. According to the principle of subsidiarity, decisions should be made at the lowest level possible and the highest level necessary. Subsidiarity is an effort at balancing the many necessary levels of society – and at its best, the principle of subsidiarity navigates the allocation of resources by higher levels of society to support engagement and decision making by the lower levels. Subsidiarity works best when citizens are engaged in their local communities. Examples: Participate in local government. Support local small businesses. Give local governments the most influence. State or federal governments should work to support local governments, but should intervene when necessary, for example, when a broader perspective is required or more high-level management is necessary.'' There is an icon of three hands in different colors facing each other in a pinwheel shape. Next it says, ``Care of God's Creation. We show our respect for the Creator by our stewardship of creation. Care for the earth is not just an Earth Day slogan, it is a requirement of our faith. We are called to protect people and the planet, living our faith in relationship with all of God’s creation. This environmental challenge has fundamental moral and ethical dimensions that cannot be ignored. Examples: Steward natural resources. Reject over-consumption. Spend time in nature to cultivate wonder for the natural world.'' There is an icon of a tree.
}
\end{figure}

Figures \ref{fig:infographic-p-1} and \ref{fig:infographic-p-2} show the full CST infographic provided to participants in the interviews.

\section{Additional Themes}
\begin{figure}
    \centering
    \includegraphics[width=\linewidth]{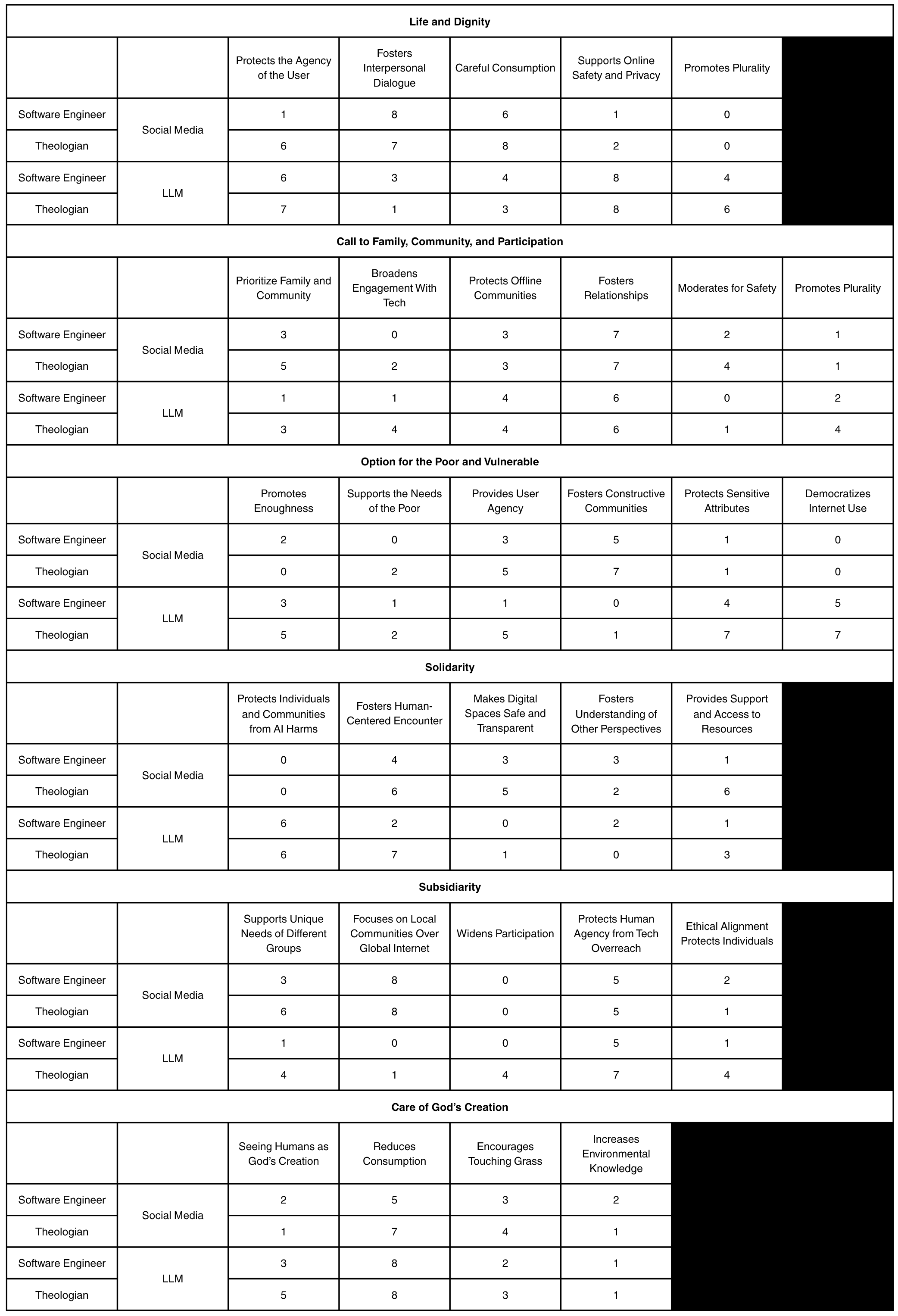}
    \caption{All themes mentioned as justifications for selecting design patterns, with the counts of how many software engineers and theologians mentioned that theme in the contexts of the social media and LLM design patterns.}
    \label{fig:all-themes}
    \Description{Comprehensive set of tables displaying participant counts across Software Engineers and Theologians for Social Media and LLM contexts, categorized under six Catholic Social Teaching principles and their sub-themes.
    - Life and Dignity:
    * Software Engineer (Social Media): Protects the Agency of the User 1; Fosters Interpersonal Dialogue 8; Careful Consumption 6; Supports Online Safety and Privacy 1; Promotes Plurality 0.
    * Theologian (Social Media): Protects the Agency of the User 6; Fosters Interpersonal Dialogue 7; Careful Consumption 8; Supports Online Safety and Privacy 2; Promotes Plurality 0.
    * Software Engineer (LLM): Protects the Agency of the User 6; Fosters Interpersonal Dialogue 3; Careful Consumption 4; Supports Online Safety and Privacy 8; Promotes Plurality 4.
    * Theologian (LLM): Protects the Agency of the User 7; Fosters Interpersonal Dialogue 1; Careful Consumption 3; Supports Online Safety and Privacy 8; Promotes Plurality 6.
    
    - Call to Family, Community, and Participation:
    * Software Engineer (Social Media): Prioritize Family and Community 3; Broadens Engagement With Tech 0; Protects Offline Communities 3; Fosters Relationships 7; Moderates for Safety 2; Promotes Plurality 1.
    * Theologian (Social Media): Prioritize Family and Community 5; Broadens Engagement With Tech 2; Protects Offline Communities 3; Fosters Relationships 7; Moderates for Safety 4; Promotes Plurality 1.
    * Software Engineer (LLM): Prioritize Family and Community 1; Broadens Engagement With Tech 1; Protects Offline Communities 4; Fosters Relationships 6; Moderates for Safety 0; Promotes Plurality 2.
    * Theologian (LLM): Prioritize Family and Community 3; Broadens Engagement With Tech 4; Protects Offline Communities 4; Fosters Relationships 6; Moderates for Safety 1; Promotes Plurality 4.
    
    - Option for the Poor and Vulnerable:
    * Software Engineer (Social Media): Promotes Enoughness 2; Supports the Needs of the Poor 0; Provides User Agency 3; Fosters Constructive Communities 5; Protects Sensitive Attributes 1; Democratizes Internet Use 0.
    * Theologian (Social Media): Promotes Enoughness 0; Supports the Needs of the Poor 2; Provides User Agency 5; Fosters Constructive Communities 7; Protects Sensitive Attributes 1; Democratizes Internet Use 0.
    * Software Engineer (LLM): Promotes Enoughness 3; Supports the Needs of the Poor 1; Provides User Agency 1; Fosters Constructive Communities 0; Protects Sensitive Attributes 4; Democratizes Internet Use 5.
    * Theologian (LLM): Promotes Enoughness 5; Supports the Needs of the Poor 2; Provides User Agency 5; Fosters Constructive Communities 1; Protects Sensitive Attributes 7; Democratizes Internet Use 7.
    
    - Solidarity:
    * Software Engineer (Social Media): Protects Individuals and Communities from AI Harms 0; Fosters Human-Centered Encounter 4; Makes Digital Spaces Safe and Transparent 3; Fosters Understanding of Other Perspectives 3; Provides Support and Access to Resources 1.
    * Theologian (Social Media): Protects Individuals and Communities from AI Harms 0; Fosters Human-Centered Encounter 6; Makes Digital Spaces Safe and Transparent 5; Fosters Understanding of Other Perspectives 2; Provides Support and Access to Resources 6.
    * Software Engineer (LLM): Protects Individuals and Communities from AI Harms 6; Fosters Human-Centered Encounter 2; Makes Digital Spaces Safe and Transparent 0; Fosters Understanding of Other Perspectives 2; Provides Support and Access to Resources 1.
    * Theologian (LLM): Protects Individuals and Communities from AI Harms 6; Fosters Human-Centered Encounter 7; Makes Digital Spaces Safe and Transparent 1; Fosters Understanding of Other Perspectives 0; Provides Support and Access to Resources 3.
    
    - Subsidiarity:
    * Software Engineer (Social Media): Supports Unique Needs of Different Groups 3; Focuses on Local Communities Over Global Internet 8; Widens Participation 0; Protects Human Agency from Tech Overreach 5; Ethical Alignment Protects Individuals 2.
    * Theologian (Social Media): Supports Unique Needs of Different Groups 6; Focuses on Local Communities Over Global Internet 8; Widens Participation 0; Protects Human Agency from Tech Overreach 5; Ethical Alignment Protects Individuals 1.
    * Software Engineer (LLM): Supports Unique Needs of Different Groups 1; Focuses on Local Communities Over Global Internet 0; Widens Participation 0; Protects Human Agency from Tech Overreach 5; Ethical Alignment Protects Individuals 1.
    * Theologian (LLM): Supports Unique Needs of Different Groups 4; Focuses on Local Communities Over Global Internet 1; Widens Participation 4; Protects Human Agency from Tech Overreach 7; Ethical Alignment Protects Individuals 4.
    
    - Care of God's Creation:
    * Software Engineer (Social Media): Seeing Humans as God's Creation 2; Reduces Consumption 5; Encourages Touching Grass 3; Increases Environmental Knowledge 2.
    * Theologian (Social Media): Seeing Humans as God's Creation 1; Reduces Consumption 7; Encourages Touching Grass 4; Increases Environmental Knowledge 1.
    * Software Engineer (LLM): Seeing Humans as God's Creation 3; Reduces Consumption 8; Encourages Touching Grass 2; Increases Environmental Knowledge 1.
    * Theologian (LLM): Seeing Humans as God's Creation 5; Reduces Consumption 8; Encourages Touching Grass 3; Increases Environmental Knowledge 1.}
\end{figure}

Figure \ref{fig:all-themes} describes all justification themes with counts of how many software engineers and theologians mentioned that theme.

The themes that had neither major similarity nor major difference in responses between the theologian and software engineer groups (i.e. the themes not described in the main paper) are described here.

\subsubsection{Life and Dignity of the Human Person}

\textbf{Promotes plurality} was only mentioned in the case of LLM chatbots, and refers to chatbots aligning to different communities' needs. This includes the chatbots being available in multiple languages: \textit{``promoting the accuracy of these systems in working in multiple languages, not just English, which is a dominant language...it's empowering to people to be able to work in their own language''} (T5). Another element of this theme is decentralizing control from tech companies. As T1 explained, \textit{``the people who are controlling [AI] have a lot of power over others and they use it to gain power...[rights reinforcement] seems like one method of trying to equalize that playing field.''}

\subsubsection{Call to Family, Community, and Participation}

\textbf{Prioritizes family and community} refers to features of social media platforms that build community, especially but not limited to supporting relationships within families and closer connections. One design features that enables this is sharing/communicating only with closer connections: \textit{``not everything you do...should be shared with everybody...there's this distinction between a more immediate community and a wider...community} (T1). Another feature is groups that are bounded based on local area or a common interest: \textit{``communities are usually bounded. You have to have boundaries in order to gain a certain type of cohesion''} (T5). For LLMs this entails avoiding humanlike chatbots that could trick users into feeling like the chatbot is family or a friend. T1 explained this with a real-life example: \textit{``this [chatbot] is not family...I have a teenager who I know is asking life questions to Google...my advice to him has been, this is not a person that cares about you. If it's a question you don't want to ask your own parents, ask someone at church, ask someone that who's an older adult that you respect...in that sense, anti-anthropomorphic AI can actually serve the call to family and community because it helps us, especially young people, distinguish between human...advice and...mechanic LLMs.''}

\textbf{Protects offline communities} refers to technologies that promote rather than distract from in-person relationships, social media recommendation algorithms that show content about one's local community, and recommendation algorithms and LLM chatbot responses that don't undermine democracy. SE1 described a relevant experience in a Discord server: \textit{``seeing those groups being used to...tell everyone here's something that I care about...even as simple as, my brother's working on this piece of art that I'd like you all to go see, or my dad's having a concert...those kind of community building events tend to happen in small groups and not on sort of feed-based, fame-based media.''} Multiple participants also mentioned that fewer notifications can help protect offline communities: \textit{``less notifications means more time with your friends and family, because you're not distracted by...your phone''} (T8). LLM systems that are less human-like may also promote less time online. SE3 said about the anti-anthropomorphic LLM design pattern, \textit{``you want to encourage human connection and...participating in the community instead of just sitting at home and talking to an AI.''}

\textbf{Moderates for safety} refers to moderated online communities and content moderation on social media platforms that help keep users, especially vulnerable users, safe. T8 mentioned in the context of moderation, \textit{``we want to have discussions about politics, but we do want to keep it...moderated...it's not helpful [when] you go into some news story’s comment section [and] it's like, bleepity bleep bleep bleep.''} One participant also referred to AI alignment in relation to this theme: \textit{``ethics and safety, that's important also to protect the family...to help create a whole ethical environment for communities in general''} (T4).

\textbf{Promotes plurality} refers to the ability for diverse communities to form on social media platforms especially through self-moderated communities, the ability to learn about other people and communities through content on social media, and LLM-enabled translation allowing people from different backgrounds to participate in community with each other. T5 drew a connection between self-moderated communities and participation, observing that \textit{``the shape that the community takes is not a matter of following predetermined rules that are handed down from on high, but rather where the members of the community actually participate in shaping the terms of conversation.''} In terms of learning about other communities on social media, SE3 said about the clear algorithmic comprehension design pattern, \textit{``maybe it can push people more content on the importance of family, community, and participation, instead of just commercials and trying to make you buy more things...[it] can incorporate recommendations on different communities and people of different religions...if you want to be more visible or include more people.''} Finally, enabling LLMs in more languages can both \textit{``broaden the community that's participating''} (T4) and \textit{``[enable communication with] somebody who you don't share a common language with... increas[ing] the bonds of unity across the human family''} (T3).

\subsubsection{Option for the Poor and Vulnerable}
The overarching themes for why the design patterns embodied the \textit{option for the poor and vulnerable} were \textbf{promotes enoughness}, \textbf{supports the needs  of the poor}, \textbf{provides user agency}, \textbf{fosters constructive communities}, \textbf{protects sensitive attributes}, and \textbf{democratizes internet use}.

\textbf{Promotes enoughness} refers to social media platforms that don't encourage consumerism and LLM technologies that limit their impact on the environment \emph{because} environmental harms disproportionately impact the poor and vulnerable. SE4 explained, \textit{``all the infrastructure we're building for these [LLM] systems is very much disproportionately hurting the poor and vulnerable. The data centers, the resource usage, electricity, water, the proximity of data centers to less affluent neighborhoods, the noise pollution they create, the heat they create...the best way to solve that is to build systems that more responsibly use resources.''} In terms of consumerism, participants saw clear algorithmic comprehension relating to the \textit{option for the poor and vulnerable} because \textit{``everyone keeps seeing ads saying, buy this, buy that...if you can control what you see, maybe then you don't see as much things and you don't feel the need to buy, and you just don't sink even deeper into the into the hole''} (SE7). 

\textbf{Supports the needs of the poor} refers to social media platforms (for example neighborhood groups) that allow the poor and vulnerable to connect with people who can help them meet their material needs and LLM chatbot responses that can point the poor and vulnerable to helpful resources. SE3 gave this LLM example: \textit{``there could be a lot of resources around them...maybe food or shelter...if someone is asking AI for help...maybe the AI can remind them they have these [resources].''} This theme also pertains to social media content that helps support the vulnerable (for example content about managing medical conditions), content from social media and LLM responses that help us better understand the needs of the poor and vulnerable, and technologies that don't distract us from the poor and vulnerable in our communities. T3 gave the example of how human-like AI can distract us from the most vulnerable in our communities: \textit{``anything in our society that's distracting has the effect of removing our attention from the suffering of the poor...we're constantly being offered opportunities to use technology and consumption to soothe ourselves, and it has the effect, I think, of making us very complacent.''}

\textbf{Fosters constructive communities} was primarily mentioned in the case of social media. This refers to platform designs that promote interpersonal communication and relationship-building (as opposed to impersonal scrolling). In relation to the friends over followers design pattern SE2 said, \textit{``the followers [have] more of the influencer dynamic, telling you something, but the friends would be someone that could ground to the real world and actually make you happier.''} This also refers to online communities that are supportive and safe, which is often achieved through moderation. SE8 noted that, \textit{``to have any sort of productive discussion, someone needs to moderate things, otherwise disagreements can get out of hand.''} The one time this theme was mentioned in the context of LLMs was about aligning LLMs to limit harmful responses: \textit{``AI of course will make it easier for all sorts of criminals to take advantage of people''} (T6).

\textbf{Democratizes internet use} was only mentioned in the case of LLMs, and refers to increased access in non-English languages and reducing the costs of using AI so that more people can benefit from it. SE7 higlighted the importance of having LLMs work well in other languages: \textit{``most of the folks that are poor and vulnerable usually don't speak English...I think that 5 years [from now] that whoever doesn't have [AI access] will be at a disadvantage.''} SE2 found that the careful context design pattern could be a way to make AI more affordable for more people: \textit{``if you're making these models more accessible by not having...people with way more resources...consuming all the tokens...this would actually democratize the AI and democratize the costs more.''}

\subsubsection{Solidarity}

\textbf{Makes digital spaces safe and transparent} refers to social media designs (especially moderation) that encourage respectful conduct, and technology designs that limit the propagation of harmful content. On moderation SE5 said, \textit{``there's always going to be people that are going to go too far, or shouldn't be allowed in the group in the first place because they're not going to allow for productive conversation to happen.''} Related to harmful content, T2 said, \textit{``[if clear algorithmic comprehension is not used then] the algorithms, for instance, might feed me things that don't bring me to solidarity, but bring me to anger and fear.''}

\textbf{Fosters understanding of other perspectives} refers to technology designs that lead to increased understanding among people with differing views, whether that's conversation-based social media platforms or LLM chatbots surfacing various perspectives: \textit{``you would want your LLMs to challenge your ideas and promote things you hadn't thought of, whether that's... about this group of people from another country, or...a group of people less fortunate than you, or maybe your ideas are excluding this thought''} (SE1). It also refers to social media feeds that recommend unbiased and accurate news as opposed to promoting \textit{``echo chambers...because extreme content gets fused''} (T3).

\subsubsection{Subsidiarity}

\textbf{Protects human agency from tech overreach} refers to technology designs that respect personal agency, especially through the ability to customize LLM parameters, control algorithmic recommendations, and limit attention-grabbing mechanisms. T7 explained that \textit{``the attention-related [design patterns], so the notification and the algorithmic comprehension...give the user more control over the system and those decisions aren't just being made by the top.''}

\subsubsection{Care of God's Creation}

\textbf{Encourages touching grass} refers to technology designs that limit our time and attention online, making us more free to encounter the people and environment around us. T4 explained, \textit{``for care of creation, it's very important to be attentive to creation, right? And sometimes social media can really distract us and make us waste time, attention, et cetera.''}

\textbf{Increases environmental knowledge} refers to technologies that allow us to increase our knowledge of environmentalism. The includes discussion forums about environmental concerns: T3 mentioned she \textit{``had great experiences with tree care subreddits''}. Other participants mentioned recommendation algorithms that promote information about caring for the environment. SE3 thought recommendation algorithms that encourage \textit{``people to enjoy nature instead of just [staying] at home''} as well as recommendation algorithms that promote more eco-friendly products would promote care of creation. This could also entail LLM responses that support the environment, such as trip-planning responses that don't \textit{``tell me to drive in places I know that I don't need to drive in''} (SE8).

\textbf{Seeing humans as God's creation} refers to the human dimension of \textit{care of God's creation}. Many of our theologian participants noted this using theological verbiage, for example, \textit{``I  think the anti-anthropomorphic AI would contribute to [care of God's creation], insofar as we recognize the distinctiveness of created human persons, and our specific role in God's creation, and just the difference between a created being and an AI tool...recognizing kind of those ontological differences, it's a necessary aspect of creation care''} (T7). Our software engineer participants intuited this connection between the distinctiveness of persons and care of God's creation without using theological language. For example, \textit{``chats are a better mechanism for connecting people and creating respect between people, and this shows...love for God's creations''} (SE2). Participants also mentioned technologies that respect human rights as upholding care of God's creation: \textit{``as human beings, we all have rights that should be guaranteed, from how our data is used to train AI, or how it's collected, how it is saved in the storage to how we train the model to make sure the model doesn't say crazy things''} (SE3).

\subsection{Fishers Exact Test Figures}
\begin{figure}
    \centering
    \includegraphics[width=\linewidth]{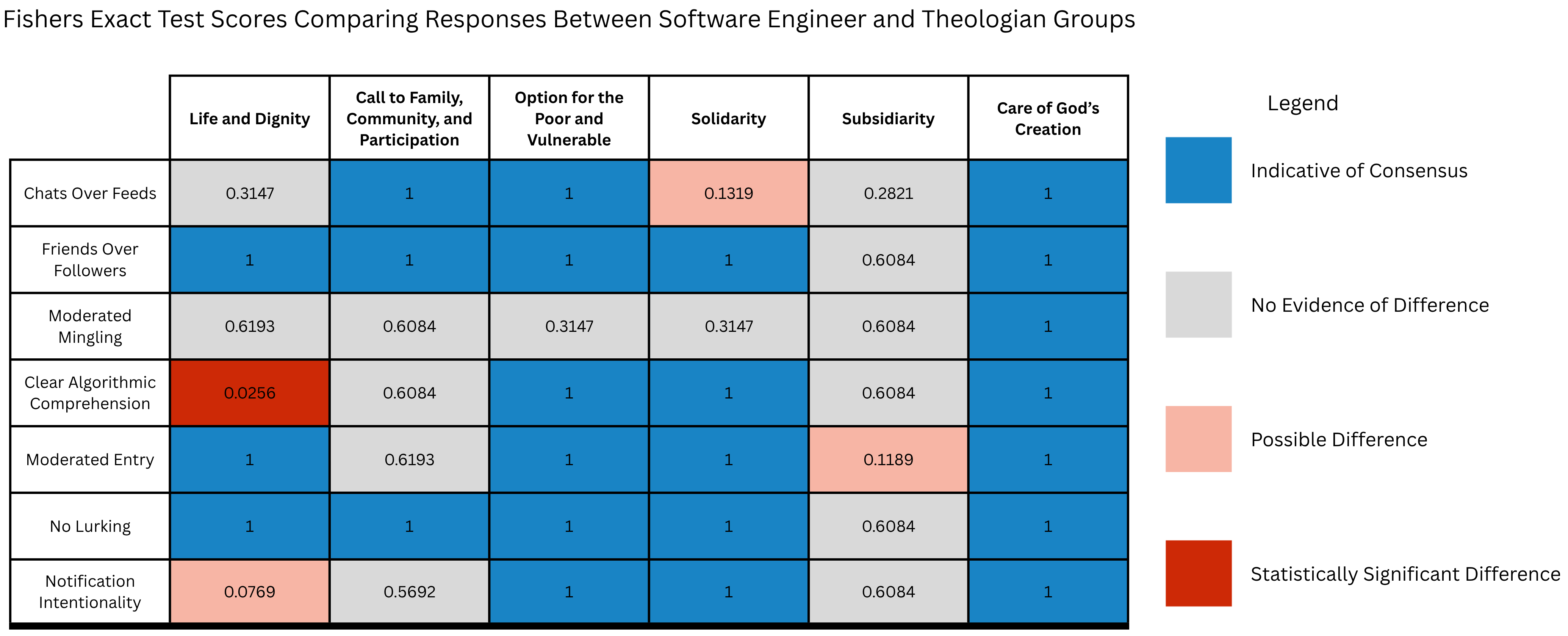}
    \caption{Results of the Fishers Exact Test comparing responses between the theologian and software engineer groups for the social media design patterns. We highlight statistically significant differences (p < 0.05), suggestion of difference (p < 0.2), no evidence of difference (0.2 <= p < 1), and indication of consensus (p = 1). For the most part there was no evidence of difference or indication of consensus between groups. However, \textit{life and dignity of the human person} had the strongest indication of difference between groups with \textit{solidarity} and \textit{subsidiarity} also suggesting possible differences.}
    \label{fig:sm-fishers}
    \Description{Table titled "Fisher's Exact Test Scores Comparing Responses Between Software Engineer and Theologian Groups", displaying p-values across seven design patterns and six CST principles with color-coded statistical significance levels:
    - Chats Over Feeds: Life and Dignity (0.3147, No Evidence of Difference); Call to Family (1, Indicative of Consensus); Option for Poor and Vulnerable (1, Indicative of Consensus); Solidarity (0.1319, Possible Difference); Subsidiarity (0.2821, No Evidence of Difference); Care of God's Creation (1, Indicative of Consensus).
    - Friends Over Followers: Life and Dignity (1, Indicative of Consensus); Call to Family (1, Indicative of Consensus); Option for Poor and Vulnerable (1, Indicative of Consensus); Solidarity (1, Indicative of Consensus); Subsidiarity (0.6084, No Evidence of Difference); Care of God's Creation (1, Indicative of Consensus).
    - Moderated Mingling: Life and Dignity (0.6193, No Evidence of Difference); Call to Family (0.6084, No Evidence of Difference); Option for Poor and Vulnerable (0.3147, No Evidence of Difference); Solidarity (0.3147, No Evidence of Difference); Subsidiarity (0.6084, No Evidence of Difference); Care of God's Creation (1, Indicative of Consensus).
    - Clear Algorithmic Comprehension: Life and Dignity (0.0256, Statistically Significant Difference); Call to Family (0.6084, No Evidence of Difference); Option for Poor and Vulnerable (1, Indicative of Consensus); Solidarity (1, Indicative of Consensus); Subsidiarity (0.6084, No Evidence of Difference); Care of God's Creation (1, Indicative of Consensus).
    - Moderated Entry: Life and Dignity (1, Indicative of Consensus); Call to Family (0.6193, No Evidence of Difference); Option for Poor and Vulnerable (1, Indicative of Consensus); Solidarity (1, Indicative of Consensus); Subsidiarity (0.1189, Possible Difference); Care of God's Creation (1, Indicative of Consensus).
    - No Lurking: Life and Dignity (1, Indicative of Consensus); Call to Family (1, Indicative of Consensus); Option for Poor and Vulnerable (1, Indicative of Consensus); Solidarity (1, Indicative of Consensus); Subsidiarity (0.6084, No Evidence of Difference); Care of God's Creation (1, Indicative of Consensus).
    - Notification Intentionality: Life and Dignity (0.0769, Possible Difference); Call to Family (0.5692, No Evidence of Difference); Option for Poor and Vulnerable (1, Indicative of Consensus); Solidarity (1, Indicative of Consensus); Subsidiarity (0.6084, No Evidence of Difference); Care of God's Creation (1, Indicative of Consensus).}
\end{figure}

\begin{figure}
    \centering
    \includegraphics[width=\linewidth]{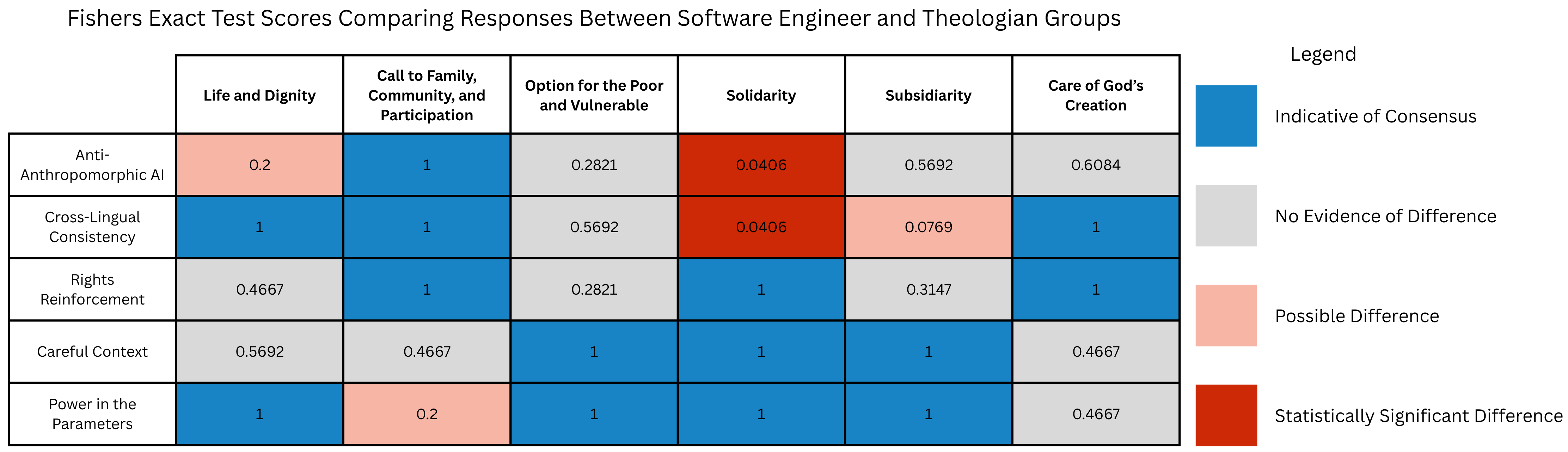}
    \caption{Results of the Fishers Exact Test comparing responses between the theologian and software engineer groups for the LLM design patterns. We highlight statistically significant differences (p < 0.05), suggestion of difference (p < 0.2), no evidence of difference (0.2 <= p < 1), and indication of consensus (p = 1). For the most part there was no evidence of difference or indication of consensus between groups. However, solidarity showed some statistically significant differences, and \textit{life and dignity}, \textit{call to family}, and \textit{subsidiarity} showed possible difference.}
    \label{fig:ai-fishers}
    \Description{Table titled "Fishers Exact Test Scores Comparing Responses Between Software Engineer and Theologian Groups", displaying p-values across five LLM design patterns and six CST principles with color-coded statistical significance levels:
    - Anti-Anthropomorphic AI: Life and Dignity (0.2, Possible Difference); Call to Family, Community, and Participation (1, Indicative of Consensus); Option for the Poor and Vulnerable (0.2821, No Evidence of Difference); Solidarity (0.0406, Statistically Significant Difference); Subsidiarity (0.5692, No Evidence of Difference); Care of God's Creation (0.6084, No Evidence of Difference).
    - Cross-Lingual Consistency: Life and Dignity (1, Indicative of Consensus); Call to Family, Community, and Participation (1, Indicative of Consensus); Option for the Poor and Vulnerable (0.5692, No Evidence of Difference); Solidarity (0.0406, Statistically Significant Difference); Subsidiarity (0.0769, Possible Difference); Care of God's Creation (1, Indicative of Consensus).
    - Rights Reinforcement: Life and Dignity (0.4667, No Evidence of Difference); Call to Family, Community, and Participation (1, Indicative of Consensus); Option for the Poor and Vulnerable (0.2821, No Evidence of Difference); Solidarity (1, Indicative of Consensus); Subsidiarity (0.3147, No Evidence of Difference); Care of God's Creation (1, Indicative of Consensus).
    - Careful Context: Life and Dignity (0.5692, No Evidence of Difference); Call to Family, Community, and Participation (0.4667, No Evidence of Difference); Option for the Poor and Vulnerable (1, Indicative of Consensus); Solidarity (1, Indicative of Consensus); Subsidiarity (1, Indicative of Consensus); Care of God's Creation (0.4667, No Evidence of Difference).
    - Power in the Parameters: Life and Dignity (1, Indicative of Consensus); Call to Family, Community, and Participation (0.2, Possible Difference); Option for the Poor and Vulnerable (1, Indicative of Consensus); Solidarity (1, Indicative of Consensus); Subsidiarity (1, Indicative of Consensus); Care of God's Creation (0.4667, No Evidence of Difference).}
\end{figure}

Figures \ref{fig:sm-fishers} and \ref{fig:ai-fishers} show the entirety of the results of the Fishers Exact Test.

\end{document}